\documentclass[
  aps,
  prd,                 
  twocolumn,           
  superscriptaddress,  
  nofootinbib,         
  amsmath,amssymb,
  longbibliography,     
]{revtex4-2}

\usepackage{graphicx}
\usepackage{dcolumn}
\usepackage{bm}
\usepackage{amsthm}
\usepackage{mathrsfs} 
\usepackage{tikz}
\usetikzlibrary{positioning, arrows.meta, calc}
\usepackage{xcolor}
\usepackage[utf8]{inputenc}
\usepackage[T1]{fontenc}
\usepackage{etoolbox}
\usepackage{hyperref}
\hypersetup{
  colorlinks=true,
  linkcolor=blue,
  citecolor=blue,
  urlcolor=blue
}
\usepackage{orcidlink}

\DeclareMathAlphabet{\mathcal}{OMS}{cmsy}{m}{n}

\theoremstyle{plain} 

\theoremstyle{definition} 

\begin{document}

\preprint{APS/PRD-2025-DRAFT}

\title{Gauge Theoretic Signal Processing II: Zero-Latency Whitening for Early Warning Pipelines}

\author{James Kennington \orcidlink{0000-0002-6899-3833}}
\email{james.kennington@ligo.org}
\affiliation{Department of Physics, The Pennsylvania State University, University Park, PA 16802, USA}
\affiliation{Institute for Gravitation and the Cosmos, The Pennsylvania State University, University Park, PA 16802, USA}

\author{Joshua Black \orcidlink{0009-0002-3746-408X}}
\affiliation{Department of Physics, The Pennsylvania State University, University Park, PA 16802, USA}
\affiliation{Institute for Gravitation and the Cosmos, The Pennsylvania State University, University Park, PA 16802, USA}

\author{Zach Yarbrough \orcidlink{0000-0002-9825-1136}}
\affiliation{Department of Physics and Astronomy, Louisiana State University, Baton Rouge, LA 70803, USA}

\author{Yun-Jing Huang \orcidlink{0000-0002-2952-8429}}
\affiliation{Department of Physics, The Pennsylvania State University, University Park, PA 16802, USA}
\affiliation{Institute for Gravitation and the Cosmos, The Pennsylvania State University, University Park, PA 16802, USA}

\author{Chad Hanna \orcidlink{0000-0002-0965-7493}}
\affiliation{Department of Physics, The Pennsylvania State University, University Park, PA 16802, USA}
\affiliation{Institute for Gravitation and the Cosmos, The Pennsylvania State University, University Park, PA 16802, USA}
\affiliation{Department of Astronomy and Astrophysics, The Pennsylvania State University, University Park, PA 16802, USA}
\affiliation{Institute for Computational and Data Sciences, The Pennsylvania State University, University Park, PA 16802, USA}

\author{Leo Tsukada \orcidlink{0000-0003-0596-5648}}
\affiliation{Department of Physics and Astronomy, University of Nevada, Las Vegas, 4505 South Maryland Parkway, Las Vegas, NV 89154, USA}
\affiliation{Nevada Center for Astrophysics, University of Nevada, Las Vegas, Las Vegas, NV 89154, USA}

\author{Amanda Baylor \orcidlink{0000-0003-0918-0864}}
\affiliation{Leonard E. Parker Center for Gravitation, Cosmology, and Astrophysics, University of Wisconsin-Milwaukee, Milwaukee, WI 53201, USA}

\author{Olivia Godwin \orcidlink{0000-0002-7489-4751}}
\affiliation{LIGO Laboratory, California Institute of Technology, MS 100-36, Pasadena, California 91125, USA}

\author{Prathamesh Joshi \orcidlink{0000-0002-4148-4932}}
\affiliation{School of Physics, Georgia Institute of Technology, Atlanta, GA 30332, USA}

\author{Cody Messick \orcidlink{0000-0002-8230-3309}}
\affiliation{Leonard E. Parker Center for Gravitation, Cosmology, and Astrophysics, University of Wisconsin-Milwaukee, Milwaukee, WI 53201, USA}

\author{Surabhi Sachdev \orcidlink{0000-0002-0525-2317}}
\affiliation{School of Physics, Georgia Institute of Technology, Atlanta, GA 30332, USA}

\author{Ron Tapia} 
\affiliation{Department of Physics, The Pennsylvania State University, University Park, PA 16802, USA}
\affiliation{Institute for Computational and Data Sciences, The Pennsylvania State University, University Park, PA 16802, USA}

\date{\today}

\begin{abstract}
Low-latency gravitational-wave search pipelines are essential for providing early warning alerts of multimessenger astrophysical transients.
Current pipelines whiten the incoming data stream using acausal, linear-phase filters, which require a look-ahead data buffer that introduces several seconds of algorithmic latency.
Eliminating this latency requires transitioning to causal, minimum-phase whitening filters that use only past data.
However, operating causal filters under non-stationary noise requires more than simply substituting the filter type: the drifting power spectral density must be tracked without degrading the matched-filter signal-to-noise ratio, filter updates must preserve the minimum-phase condition, and the altered phase response must be compensated to maintain sky localization accuracy.
In Paper~I of this series, we introduced a gauge theoretic signal processing framework and showed that the minimum-phase connection on the manifold of power spectra provides a geometrically exact update rule for causal filters.
In this paper, we validate that framework numerically and operationally, demonstrating that parallel transport along this connection strictly preserves the minimum-phase property while exactly conserving the matched-filter signal-to-noise ratio.
We numerically certify the flatness of this connection, demonstrating that the optimal filter is a path-independent state function of the instantaneous noise.
Through an injection campaign on O3 data with 15,347 binary black hole signals across the LIGO-Virgo network, we confirm that this zero-look-ahead architecture preserves the detection sensitivity and inter-detector timing and phase accuracy of the standard linear-phase baseline.
Implementing this framework within the production \textsc{sgnl} matched-filter pipeline reduces the whitening latency by 1.0 second (33\%) relative to the standard linear-phase baseline at a 4-second noise estimation cadence, confirmed both in controlled local tests and on live O3 replay data at production scale.
Stride reduction experiments demonstrate that up to 91\% of the baseline trigger latency can be eliminated with sub-second pipeline cadence.
\end{abstract}

\maketitle

\section{Introduction}
\label{sec:intro}

The rapid identification of compact binary coalescences is a primary objective of the ground-based gravitational-wave observatory network~\cite{TheLIGOScientificCollaboration2015Advanced, Acernese2015Advanced, Akutsu2021Overview}.
For systems involving matter, such as binary neutron stars and neutron star-black hole binaries, extracting the gravitational-wave signal before the exact moment of merger is critical for enabling multimessenger astronomy~\cite{the_ligo_scientific_collaboration_gw170817_2017}.
Early warning alerts allow electromagnetic observatories to slew to target coordinates and capture prompt emission across the electromagnetic spectrum~\cite{nitz_gravitational-wave_2020, magee_first_2021, kovalam_early_2022}.
Consequently, the utility of a low-latency search pipeline is strictly bottlenecked by its algorithmic latency~\cite{messickAnalysisFrameworkPrompt2017b, nitz_rapid_2018, chu_spiir_2021, adams_low-latency_2016}.
Any delay between the acquisition of physical strain data and the generation of a matched-filter trigger directly reduces the time available for subsequent telescope repositioning~\cite{cannon_toward_2012}.

Optimal signal extraction via matched filtering requires precise knowledge of the detector noise floor, which is parameterized by the power spectral density (PSD)~\cite{wainstein_extraction_1970, helstrom_statistical_1968, allen_findchirp_2012}.
Ground-based interferometers are fundamentally non-stationary instruments~\cite{abbottGuideLIGODetector2020, davisLIGODetectorCharacterization2021}: the spectrum fluctuates over time due to thermal drift~\cite{saulson1990thermal}, microseismic activity~\cite{matichard2015seismic}, and transient instrumental artifacts~\cite{abbottCharacterizationTransientNoise2016, zackay_detecting_2021}.
To maintain optimal sensitivity, search pipelines must dynamically estimate the noise floor and continuously update the whitening filter applied to the incoming data stream~\cite{messickAnalysisFrameworkPrompt2017b}.
Current production pipelines accomplish this using acausal, linear-phase filters within a weighted overlap-add (WOLA) architecture~\cite{oppenheim_discrete-time_2010, sachdev_gstlal_2019, ewingPerformanceLowlatencyGstLAL2024}.
This approach is effective, it successfully tracks PSD drift while preserving detection sensitivity, but the symmetric filter inherently requires a look-ahead data buffer to compute its output, introducing a fixed algorithmic latency that scales with the PSD estimation window length~\cite{tsukada_application_2018}.
For the standard online configuration, this latency amounts to several seconds, precluding sub-second early warning targets.

Eliminating this latency bottleneck requires transitioning from acausal, linear-phase whitening to causal, minimum-phase whitening, which uses only past data and requires no look-ahead buffer~\cite{KenningtonBlack2026_GTSP1, nussenzveig1972causality}.
However, dynamically updating a causal whitening filter as the detector noise evolves introduces two challenges that do not arise in the acausal setting.
First, the set of minimum-phase filters does not form a vector space: naive linear interpolation between two valid causal filters can force transfer function roots outside the unit circle, violating the minimum-phase condition~\cite{bode1945network, oppenheim_discrete-time_2010}.
Second, any update rule must track the drifting PSD without introducing dispersive phase accumulation~\cite{toll1956causality, nussenzveig1972causality} or degrading the matched-filter signal-to-noise ratio (SNR)~\cite{helstrom_statistical_1968}, requirements that are automatically satisfied in the linear-phase case but must be explicitly enforced for causal filters.

In Part~I of this series~\cite{KenningtonBlack2026_GTSP1}, we demonstrated that both challenges are resolved by reframing the problem geometrically.
We introduced the gauge theoretic signal processing (GTSP) framework, in which the set of invertible minimum-phase filters forms a Lie group~\cite{hall2015lie, warner1983foundations, gilmore2008lie}.
We defined a holonomic connection on the manifold of logarithmic power spectra~\cite{lee1997riemannian, gockeler1989differential, baggio_conal_2018} that provides the exact parallel transport of the whitening filter along the drift path of the detector noise state.
This connection simultaneously preserves the minimum-phase property, solving the first challenge, and conserves the matched-filter SNR without dispersive phase, solving the second.

In this paper, we provide the numerical and pipeline-level validation of the GTSP framework.
In Section~\ref{sec:numerical}, we first characterize the causality violation rate of linear filter interpolation on real LIGO O4a data, establishing the practical relevance of the minimum-phase constraint.
We then numerically verify the two defining properties of this connection: its exact path independence (flatness) across $10^6$ closed loops in PSD space~\cite{spivak1999comprehensive, ambrose1953theorem}, and its machine-precision conservation of the matched-filter SNR along $10^6$ random drift paths.

In Section~\ref{sec:pipeline}, we integrate the holonomic update into the production \textsc{sgnl} matched-filter pipeline~\cite{cannon_toward_2012, messickAnalysisFrameworkPrompt2017b, huang_sgnl_2025} and evaluate its operational performance on real O3 data with 15,347 BBH injections across the LIGO-Virgo detector network.
We confirm that the zero-latency architecture preserves both the detection sensitivity and the inter-detector timing and phase accuracy required for sky localization~\cite{fairhurst_triangulation_2009, singer_rapid_2016}.
We then derive an analytic latency model showing that the causal whitening architecture achieves a constant trigger latency independent of the PSD estimation window length, verify this prediction against measured pipeline profiling, and confirm the whitening gain at production scale on live O3 replay data.

\section{Numerical Validation of the Geometric Framework}
\label{sec:numerical}
In the first paper of this series~\cite{KenningtonBlack2026_GTSP1}, we established that the space of causal spectral factors carries a principal bundle structure, within which the minimum-phase filters form a distinguished global section (see Paper~I, Section~II~\cite{baggio_conal_2018}).
To deploy this continuous mathematical framework within a discrete low-latency pipeline, we must computationally validate its discrete algorithmic implementation.
This section provides the numerical verification that bridges the GTSP formalism with the practical constraints of digital signal processing.

We structure this validation through three targeted experiments.
First, we examine the fundamental failure modes of standard adaptive methods by analyzing the topological and causal breakdown of linear filter interpolation on real LIGO O4a detector data~\cite{sayed_fundamentals_2003}.
Second, we certify the geometric consistency of the holonomic update by numerically verifying the flatness of the connection across $10^6$ closed loops in synthetic PSD space, ensuring that the filter evolution remains strictly path independent regardless of intervening transient noise~\cite{lee1997riemannian}.
Finally, we demonstrate that parallel transport along the spectral manifold exactly conserves the matched-filter detection statistic, preventing the phase and amplitude distortions traditionally introduced by discrete windowing techniques~\cite{oppenheim_discrete-time_2010}.

The flatness and SNR conservation experiments utilize synthetic power spectral densities modeled after the Advanced LIGO noise floor~\cite{buikemaSensitivityPerformanceAdvanced2020}, while the causality analysis is performed directly on two weeks of O4a science data.
The discrete filter coefficients are computed using exact cepstral projections~\cite{oppenheim_discrete-time_2010, makhoulLinearPredictionTutorial1975}, directly mirroring the operational conditions and sampling rates of current low-latency gravitational-wave searches.

\subsection{Methods: Monte Carlo Integration on the PSD Manifold}
\label{sec:numerical-methods}

To numerically verify the theoretical guarantees of the minimum-phase (MP) connection, we perform a series of Monte Carlo experiments on the manifold of power spectral densities.
We represent the PSD space non-parametrically using $N=8192$ discrete frequency bins, covering the standard LIGO observing band from 10 Hz to 2048 Hz.
This high-dimensional representation ensures that our results are not artifacts of a specific parametric model and can capture the sharp spectral features—such as mechanical resonances (violin modes) and calibration lines—that characterize modern gravitational-wave detectors~\cite{matichard2015seismic}.

The central challenge in validating a dynamical filter framework is the generation of realistic but extreme noise trajectories.
We construct these trajectories in the log-PSD domain, $\ln S(f,t)$, where the GTSP update is linear.
We employ three classes of PSD data.
The first two are synthetic: \textit{cyclic drift paths}, which are smooth, periodic trajectories constructed by superimposing $K=20$ spectral perturbation modes with time-varying coefficients, designed as closed loops in PSD space to test the holonomy (flatness) of the connection~\cite{lee1997riemannian, spivak1999comprehensive}; and \textit{random Markovian excursions}, which are stochastic paths that simulate the unmodeled drift and abrupt state changes often seen during instrument commissioning~\cite{davisLIGODetectorCharacterization2021, abbottGuideLIGODetector2020}, providing the basis for testing SNR conservation under chaotic conditions.
The third class consists of \textit{real LIGO O4a data}: two weeks of science-mode PSD snapshots (Oct 1--14, 2023) from the H1 and L1 detectors~\cite{collaborationOpenDataLIGO2025}, used to evaluate the minimum-phase violation rates of linear filter blending under operational noise conditions (Section~\ref{sec:numerical-causality}).

For the holonomy and SNR conservation tests, we integrate the filter evolution along synthetic paths using $n=500$ discrete steps.
At each step, we compare the GTSP holonomic update against alternative geometric connections (e.g., Levi-Civita, Amari $\alpha$-connections) and standard linear cross-fading baselines~\cite{amariMethodsInformationGeometry2000, sayed_fundamentals_2003}.

\subsection{Minimum-Phase Violation Under Linear Filter Blending}
\label{sec:numerical-causality}

Linear blending of whitening filter coefficients, the operation performed during a standard $\cos^2$ cross-fade between filter updates (see Appendix~\ref{app:crossfade} for a proof of this equivalence), can violate the fundamental requirement of minimum-phase~\cite{sayed_fundamentals_2003}.
This violation introduces non-causality into the filtering process, leading to timing artifacts and SNR loss~\cite{nussenzveig1972causality, toll1956causality}.

\begin{figure*}[t]
    \centering
    \includegraphics[width=\textwidth]{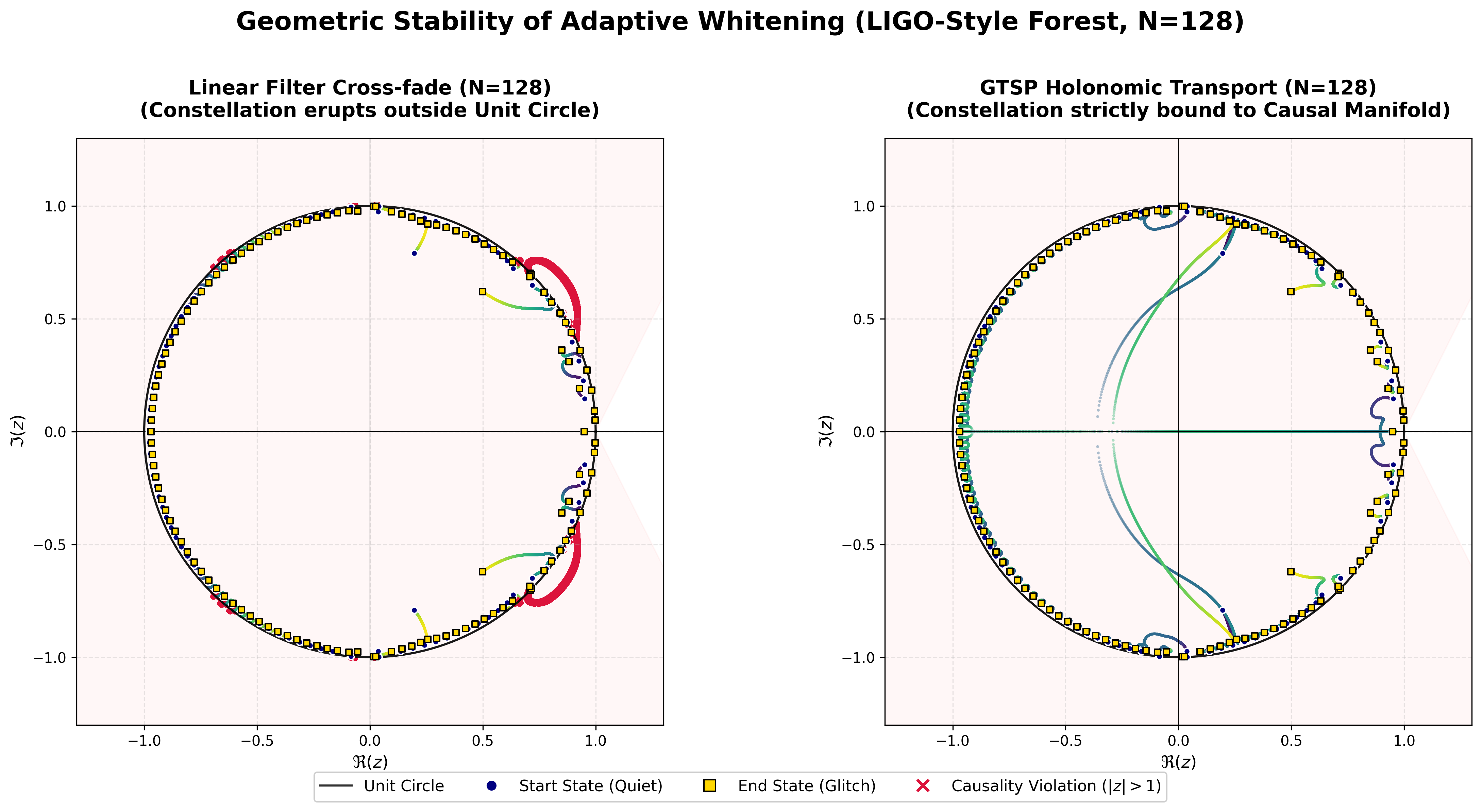}
    \caption{Z-plane root migration during a realistic LIGO PSD transition.
    (Left) Standard linear cross-fade: roots are pushed outside the unit circle, violating causality.
    (Right) GTSP holonomic transport: roots remain strictly within the unit circle, preserving the minimum-phase property by construction.}
    \label{fig:root-migration}
\end{figure*}

The failure mechanism is rooted in the fact that the set of minimum-phase filters is not a vector space.
When two minimum-phase filters \(W_A\) and \(W_B\) exhibit a phase difference approaching \(\pi\) at a specific frequency \(f_c\), their linear combination can result in a filter whose zeros in the Z-plane migrate outside the unit circle~\cite{bode1945network, oppenheim_discrete-time_2010}.
At the point in the cross-fade where the magnitudes are roughly equal, the effective filter response can vanish (\(|W_{\text{eff}}(f_c)| \approx 0\)), signaling a topological transition of the filter's root constellation.
Figure~\ref{fig:root-migration} illustrates this mechanism for a realistic LIGO PSD transition: the linear cross-fade pushes roots outside the unit circle, while the GTSP holonomic transport confines them strictly within.

Using the O4a dataset described in Section~\ref{sec:numerical-methods}, we empirically analyzed this violation rate on two weeks of real LIGO science data (Oct 1--14, 2023)~\cite{TheLIGOScientificCollaboration2015Advanced, collaborationOpenDataLIGO2025}.
For each detector (H1 and L1), we tracked the minimum-phase status of linear blends across varying time lags between PSD updates.
As shown in Table~\ref{tab:violation-rates}, the violation rate is rare but nonzero even at the standard 8-second pipeline cadence (0.017\%, or 20 violations in two weeks), and becomes significant at longer timescales.
At weekly cadences, more than half of the L1 transitions (58.2\%) result in non-causal filters.

\begin{table}[h]
\caption{Minimum-phase violation rates vs. update cadence on O4a data.}
\label{tab:violation-rates}
\begin{ruledtabular}
\begin{tabular}{lcc}
Time Lag & H1 (\%) & L1 (\%) \\
\hline
8 s (pipeline) & 0.017 & 0.017 \\
1 hour & 1.4 & 1.5 \\
12 hours & 14.3 & 13.2 \\
1 day & 14.2 & 27.5 \\
1 week & 28.6 & 58.2 \\
\end{tabular}
\end{ruledtabular}
\end{table}

The impact on gravitational-wave detection sensitivity is measurable.
While the mean SNR loss for a broadband CBC template is small (0.01\%), the worst-case loss reaches 8.0\% per transition.
This loss is particularly acute during the samples where the phase collision occurs, effectively amputating the signal's coherence for several seconds.
In contrast, the GTSP exponential-map update, by operating on the log-spectral manifold, yields zero violations and zero SNR loss across all tested cadences and data segments.
The O4a analysis establishes the operational need for minimum-phase-preserving updates; the pipeline validation on O3 data in Section~\ref{sec:pipeline} confirms the framework's viability on production detector data.

\subsection{Verification of the Flatness Theorem (Path Independence)}
\label{sec:numerical-flatness}

In Paper I~\cite{KenningtonBlack2026_GTSP1}, we established that the MP connection on the manifold of causal spectral factors is theoretically flat, ensuring that parallel transport around any closed loop in power spectral density space returns the filter exactly to its initial value.
This path-independence is the geometric property that enables bit-for-bit reproducible pipeline states across distributed, asynchronous computing clusters.
Here, we provide the numerical verification of this flatness across $10^6$ closed loops in a high-dimensional, non-parametric log-PSD space ($N=8192$)~\cite{lee1997riemannian}.

Each test loop is a smooth, band-limited trajectory constructed by superimposing $K=20$ spectral perturbation modes with periodic time-varying coefficients.
For each loop $\gamma$ ($S_0 \to S_1 \to \cdots \to S_0$), we measure the holonomy
\begin{equation}
    \mathcal{H}(\gamma) = \frac{\|W_{\text{final}} - W_{\text{initial}}\|}{\|W_{\text{initial}}\|}.
    \label{eq:holonomy}
\end{equation}
A flat connection satisfies $\mathcal{H}(\gamma) = 0$ for all loops~\cite{spivak1999comprehensive, ambrose1953theorem}.

\begin{figure*}[t]
    \centering
    \includegraphics[width=\textwidth]{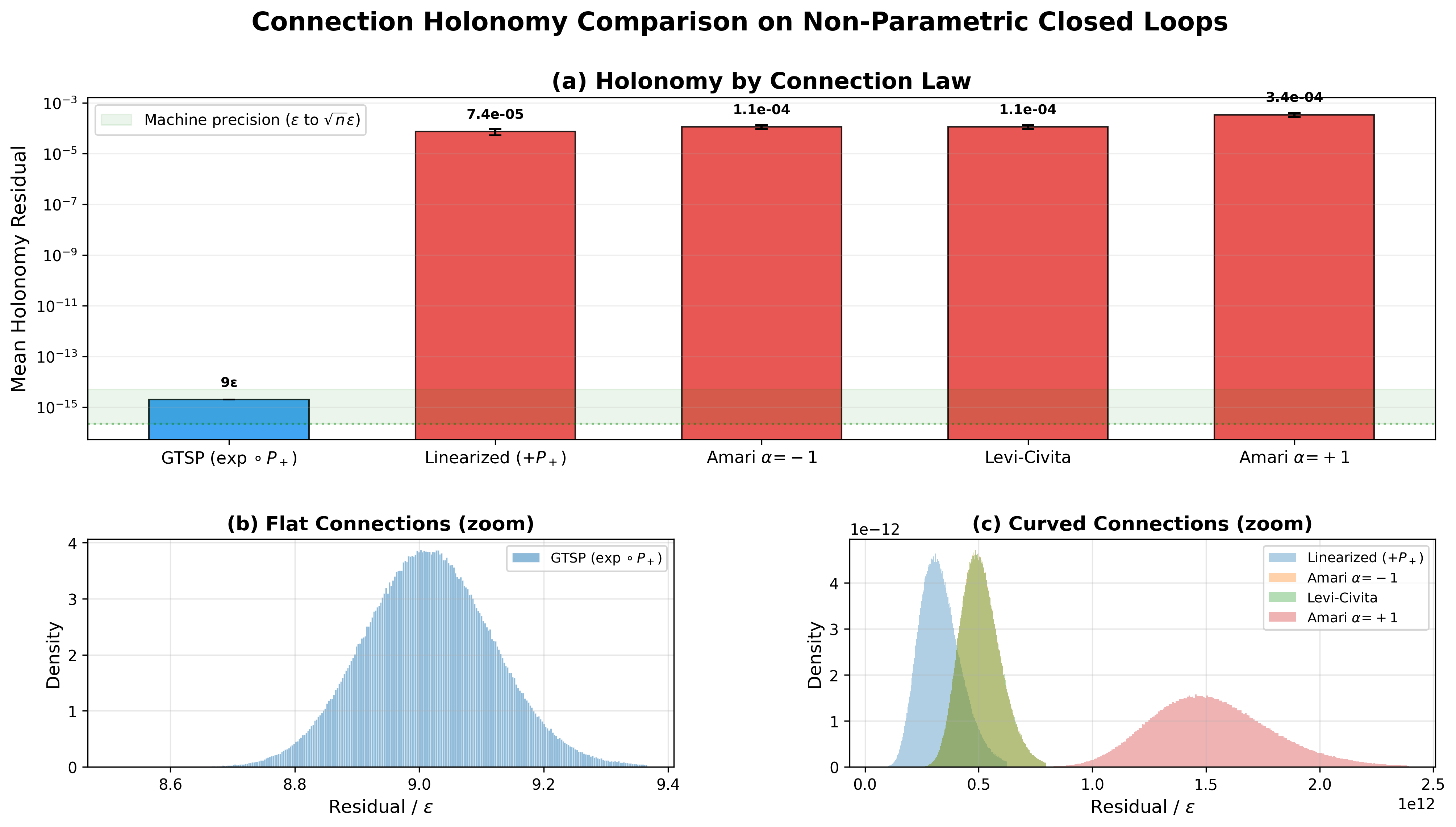}
    \caption{Numerical holonomy comparison across five connection types.
    (a) Mean holonomy on a log-scale: the MP connection (blue) remains at machine precision ($9.0\varepsilon$), while alternative geometric connections (red) exhibit curvature residuals 11 orders of magnitude higher.
    (b) Histogram of GTSP holonomy residuals in units of $\varepsilon$, showing a tight Gaussian centered at $9.0\varepsilon$.
    (c) Overlaid histograms for the curved connections (Levi-Civita, Amari m, Amari e), showing distinct curvature clustering.}
    \label{fig:holonomy-comparison}
\end{figure*}

Our results, summarized in Fig.~\ref{fig:holonomy-comparison}, confirm that the MP connection is flat to within IEEE 754 floating-point rounding error~\cite{higham2002accuracy}.
The mean holonomy of $2.00 \times 10^{-15}$ ($9.0\varepsilon$) is consistent with the expected accumulation of rounding errors over $n=500$ integration steps ($\sqrt{n}\varepsilon \approx 22\varepsilon$).

We find that flatness is uniquely possessed by the exponential-map update.
Replacing the $\exp(x)$ update with its first-order Taylor approximation $(1+x)$, effectively a linearized blend with causal projection, breaks flatness, resulting in a mean holonomy of $7.4 \times 10^{-5}$.
This confirms that the group structure of the exponential map is essential for path-independence~\cite{warner1983foundations, hall2015lie}.

Furthermore, we demonstrate that standard information-geometric connections on the PSD manifold (e.g., Levi-Civita, Amari $\alpha$-connections) exhibit significant curvature when lifted to the spectral factor bundle~\cite{amariMethodsInformationGeometry2000}.
These connections result in holonomy residuals that are $10^{11}$ to $10^{12}$ times larger than those of the MP connection.
This underscores that the minimum-phase parallel transport is not merely a valid choice but is the unique causal connection that remains flat on the spectral factor bundle~\cite{baggio_conal_2018}.

The fact that composing the exponential map with the causal projection $P_+$ does not introduce curvature is non-trivial.
The exponential connection alone is trivially flat (pointwise multiplication commutes), but $P_+$ is a non-local operation that couples all frequencies through the cepstral domain and could, in principle, break this commutativity.
It does not, because the linearity of $P_+$ ($P_+[a+b] = P_+[a] + P_+[b]$), together with the abelian structure of the scalar filter algebra, ensures that the telescoping property of the exponential updates is preserved~\cite{oppenheim_discrete-time_2010}.
This compatibility between the exponential map and the causal projection is the algebraic property underlying the flatness of the minimum-phase connection derived in Paper~I~\cite{KenningtonBlack2026_GTSP1}.

\subsection{Covariant SNR Conservation Under Parallel Transport}
\label{sec:numerical-snr}

The final numerical validation of the GTSP framework addresses its metric compatibility—specifically, the conservation of the matched-filter signal-to-noise ratio (SNR) under dynamic noise conditions.
In Paper I~\cite{KenningtonBlack2026_GTSP1}, we theoretically derived the metric compatibility condition
\begin{equation}
    \nabla_\gamma \rho = 0,
    \label{eq:snr-conservation}
\end{equation}
asserting that the optimal SNR is a covariant constant along the minimum-phase parallel transport.
We verify this claim numerically by tracking the SNR of a standard Newtonian chirp template along 1,000,000 random PSD evolution paths, each consisting of 500 integration steps~\cite{helstrom_statistical_1968, wainstein_extraction_1970}.

\begin{figure*}[t]
    \centering
    \includegraphics[width=\textwidth]{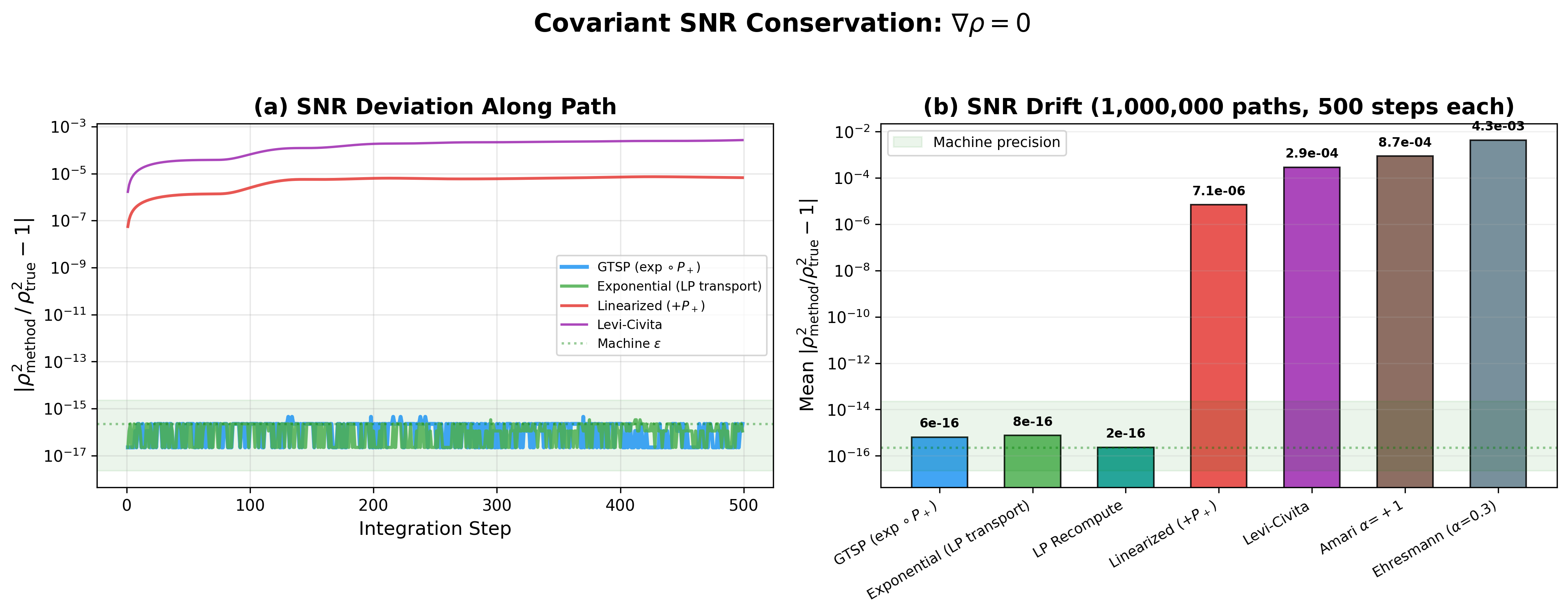}
    \caption{Covariant SNR conservation ($\nabla\rho = 0$).
    (a)~SNR deviation along a random 500-step path for different connections: GTSP and Exponential connections remain at machine precision, while the linearized blend and Ehresmann connections exhibit significant drift.
    (b)~Distribution of mean absolute SNR drift per path across 1,000,000 simulations.}
    \label{fig:snr-conservation}
\end{figure*}

Our findings, summarized in Table~\ref{tab:snr-drift}, confirm that the MP connection conserves matched-filter SNR to machine precision.
The mean SNR drift of $6.4 \times 10^{-16}$ is indistinguishable from machine epsilon ($\varepsilon$)~\cite{higham2002accuracy}, numerically validating the metric compatibility of the connection.

We compare the GTSP performance against several alternative connection laws.
While the Exponential connection (the exponential map without the causal projection $P_+$) also conserves SNR (mean drift $7.6 \times 10^{-16}$), it produces a linear-phase filter that diverges from the required minimum-phase solution, introducing dispersive group delay (mean filter error of $6.1 \times 10^{-2}$)~\cite{nussenzveig1972causality}.
Conversely, linear blending of filter coefficients (Linearized) results in a cumulative SNR drift of $7.1 \times 10^{-6}$ per step, which is ten orders of magnitude above machine precision.
This confirms that linear interpolation introduces measurable sensitivity losses, even when the transitions are smooth~\cite{sayed_fundamentals_2003}.

Furthermore, we test information-geometric connections that are geometrically valid on the PSD manifold but are not metric-compatible with the matched-filter inner product.
The Levi-Civita and Ehresmann connections exhibit significant SNR drift, with the Ehresmann connection reaching a maximum drift of 9.1\% over a single path (see Fig.~\ref{fig:snr-conservation})~\cite{amariMethodsInformationGeometry2000}.
These results verify that the MP connection is uniquely selected by the joint requirements of causality and SNR conservation.

\begin{table}[h]
\caption{Matched-filter SNR drift across $10^6$ random PSD paths.}
\label{tab:snr-drift}
\begin{ruledtabular}
\begin{tabular}{lccc}
Connection & Mean SNR drift & Max SNR drift & Filter error \\
\hline
\textbf{GTSP} & $\mathbf{6.4 \times 10^{-16}}$ & $\mathbf{1.4 \times 10^{-14}}$ & $\mathbf{2.1 \times 10^{-15}}$ \\
Exponential & $7.6 \times 10^{-16}$ & $7.2 \times 10^{-15}$ & $6.1 \times 10^{-2}$ \\
Linearized & $7.1 \times 10^{-6}$ & $8.5 \times 10^{-4}$ & $4.4 \times 10^{-4}$ \\
Levi-Civita & $2.9 \times 10^{-4}$ & $1.6 \times 10^{-3}$ & $6.1 \times 10^{-2}$ \\
Amari $m$ & $8.7 \times 10^{-4}$ & $4.8 \times 10^{-3}$ & $6.1 \times 10^{-2}$ \\
Ehresmann & $4.3 \times 10^{-3}$ & $9.1 \times 10^{-2}$ & $3.5 \times 10^{-2}$ \\
\end{tabular}
\end{ruledtabular}
\end{table}

To ensure that the minimum-phase factorization itself does not introduce sensitivity penalties, we perform a separate single-PSD parity check between acausal (linear-phase) and causal (minimum-phase) whitening.
For a stationary detector, the peak matched-filter SNR produced by both filters is identical to 14 significant digits (discrepancy of $6.7 \times 10^{-14}$ \%), with exactly zero timing offset~\cite{owen_search_1996, allen_findchirp_2012}.
This confirms that the causal factorization preserves the full theoretical sensitivity of the optimal matched filter.

\section{Pipeline Verification and Zero-Latency Implementation}
\label{sec:pipeline}
The critical question for the GTSP framework is whether the geometric properties validated numerically in the preceding section survive integration into a production gravitational-wave search pipeline, where discrete sampling, finite filter orders, and real-time throughput constraints introduce practical challenges beyond the idealized continuous theory.

This section answers that question affirmatively.
We realize the holonomic update law as a streaming adaptive filter within the production \textsc{sgnl} matched-filter pipeline~\cite{cannon_toward_2012, messickAnalysisFrameworkPrompt2017b, huang_sgnl_2025} and evaluate its performance through injection campaigns on both controlled synthetic noise and real O3 detector data.
We demonstrate three results.
First, the drift-corrected minimum-phase pipeline preserves the detection sensitivity of the linear-phase baseline across all event significance levels.
Second, the inter-detector timing and phase precision, the observables that determine sky localization accuracy, are statistically indistinguishable from those of the baseline.
Third, the causal whitening architecture eliminates the algorithmic look-ahead buffer required by standard spectral whitening, yielding a measured latency reduction that is confirmed at production scale on live O3 replay data.

\subsection{Methods: Pipeline Architecture and Injection Campaign}
\label{sec:pipeline-methods}

The numerical experiments of the preceding section validated the MP connection in the continuous limit: exact flatness, strict minimum-phase preservation, and machine-precision SNR conservation.
Paper~I derived the mathematical form of the correction kernel; here we develop its realization as a production streaming architecture, including the adjoint projection onto the data path, the stride absorption mechanism for the anti-causal kernel, and the integration with the multi-rate LLOID filterbank~\cite{cannon_toward_2012}.
Recall that the exact correction kernel mapping a reference noise state $S_{\mathrm{ref}}$ to the live state $S_{\mathrm{live}}$ is given by
\begin{equation}
    \mathcal{K} = \exp\!\left(\mathcal{P}_+\!\left[\log\frac{S_{\mathrm{ref}}}{S_{\mathrm{live}}}\right]\right),
    \label{eq:correction-kernel}
\end{equation}
where $\mathcal{P}_+$ denotes the causal (cepstral) projection~\cite{KenningtonBlack2026_GTSP1}.
Throughout this section, we denote the frequency-domain correction kernel as $\mathcal{K}$, its adjoint (complex conjugate) as $\mathcal{K}^\dagger$, and its truncated $L$-tap time-domain FIR realization as $\mathcal{K}_L$.
We integrate this update law into the \textsc{sgnl} low-latency matched-filter pipeline~\cite{cannon_toward_2012, messickAnalysisFrameworkPrompt2017b, huang_sgnl_2025}.

Central to this implementation is the \textit{adjoint projection} architecture.
As derived in Paper~I~\cite{KenningtonBlack2026_GTSP1} and detailed in Appendix~\ref{app:pipeline-topology}, the geometric correction can be deployed in three mathematically equivalent representations: a forward action on the templates, an adjoint action on the data stream, or a post-processing convolution on the SNR time series.
For a production pipeline with a massive SVD template bank, the adjoint (pullback) representation is the natural choice: the static reference templates remain unmodified in memory, while the correction kernel $\mathcal{K}^\dagger$ is applied as a streaming anti-causal finite impulse response (AFIR) filter to the incoming whitened data.
This decouples the template manifold from the noise manifold, enabling high-cadence whitening updates with $\mathcal{O}(1)$ computational cost independent of the number of templates.
The correction kernel is recomputed whenever the live PSD estimate deviates from the reference by more than a similarity threshold, typically every 8--16\,s during stable operation (see Appendix~\ref{app:pipeline-topology}).

The AFIR stage introduces a finite anti-causal delay of $(L-1)/f_s$ seconds, where $L$ is the truncation length (in taps) of the correction kernel and $f_s$ is the sample rate.
This delay is absorbed by the stride buffer of the AFIR element for $L \leq 2048$ taps (1.0\,s at $f_s = 2048$\,Hz), adding no additional wall-clock latency to the pipeline (see Sec.~\ref{sec:pipeline-latency}).

\subsubsection{Injection campaign}

We evaluate the pipeline on the same 8.8-day O3 data segment (GPS 1241725020--1242485126) used for the original \textsc{sgnl} pipeline validation~\cite{buikemaSensitivityPerformanceAdvanced2020, huang_sgnl_2025}, enabling direct comparison with the established baseline.
A total of 15,347 simulated binary black hole (BBH) signals are injected into the three-detector (H1 + L1 + V1) data stream.
The pipeline is configured identically to the production offline search: three-detector coincidence, SVD-decomposed template bank, and full matched-filter processing~\cite{messickAnalysisFrameworkPrompt2017b, ewingPerformanceLowlatencyGstLAL2024}.
The latency architecture operates entirely on the whitening path, upstream of template matching, and is therefore waveform-independent; the BBH campaign validates the sensitivity and accuracy of the whitening infrastructure under the most demanding conditions for detection (short in-band duration).

To characterize the baseline's intrinsic run-to-run variability, we execute five independent LP pipeline runs on the same data.
The LP recovery at a false alarm rate (FAR) threshold of $\leq 1\,\mathrm{yr}^{-1}$ yields a mean of 3158 injections with a standard deviation of $\sigma = 24$, reflecting the stochastic variability in PSD estimation, ranking, and FAR assignment across independent pipeline instantiations.
All GTSP configurations are compared against this variability envelope.

We compare five pipeline configurations, summarized in Table~\ref{tab:configs}.
The primary comparison is between the LP baseline and the drift-corrected GTSP architecture at $L = 128$ taps (62\,ms), the shortest truncation that captures the dominant correction, minimizing anti-causal delay while operating well within the stride absorption limit (Sec.~\ref{sec:pipeline-latency}).

\begin{table}[h]
\caption{Pipeline configurations tested in the O3 injection campaign.}
\label{tab:configs}
\begin{ruledtabular}
\begin{tabular}{lll}
Configuration & Whitening & Drift correction \\
\hline
LP baseline & Linear-phase & N/A (WOLA cross-fade) \\
Drift $L{=}128$ & Minimum-phase & $\mathcal{K}_{128}$ (62\,ms) \\
Drift $L{=}512$ & Minimum-phase & $\mathcal{K}_{512}$ (250\,ms) \\
Drift $L{=}2048$ & Minimum-phase & $\mathcal{K}_{2048}$ (1.0\,s) \\
Drift full & Minimum-phase & $\mathcal{K}_{N}$ (8.0\,s) \\
\end{tabular}
\end{ruledtabular}
\end{table}

Performance is evaluated through three complementary metrics: injection recovery rate at fixed FAR thresholds, inter-detector timing accuracy ($\Delta t_{H1-L1}$), and inter-detector phase accuracy ($\Delta\phi_{H1-L1}$).
The latter two quantities are the direct observables that determine triangulation-based sky localization~\cite{fairhurst_triangulation_2009, singer_rapid_2016}.
Statistical equivalence between configurations is assessed using Kolmogorov-Smirnov tests for distributional agreement, Levene tests for variance homogeneity, and Cohen's $d$ for effect size.

\subsection{Sensitivity Preservation}
\label{sec:pipeline-sensitivity}

We evaluate the search sensitivity of the GTSP pipeline by measuring the number of injections recovered at a fixed false alarm rate threshold of FAR~$\leq 1\,\mathrm{yr}^{-1}$ across all pipeline configurations.

\begin{figure}[t]
    \centering
    \includegraphics[width=\columnwidth]{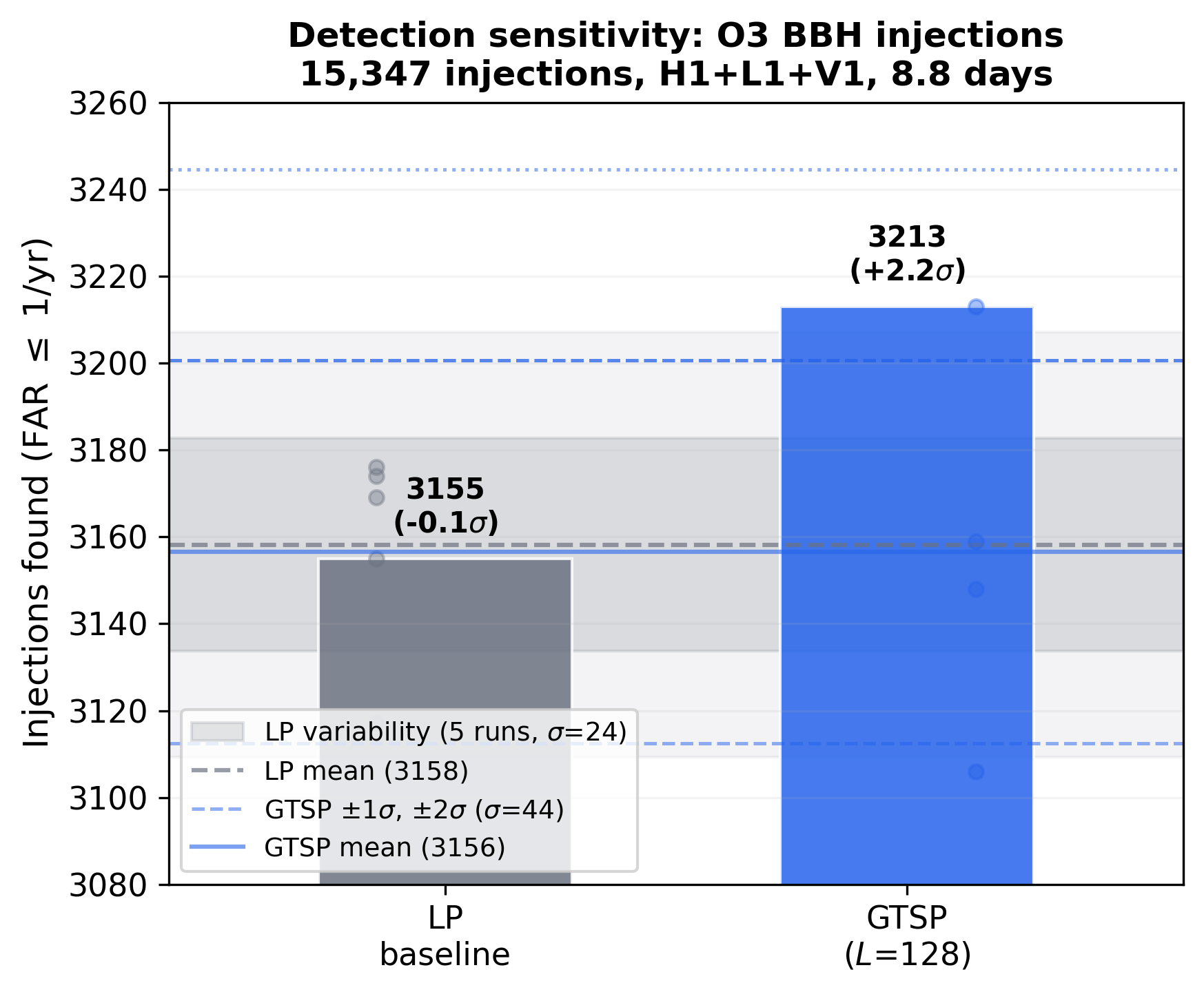}
    \caption{Detection sensitivity on O3 data with 15,347 BBH injections.
    The GTSP architecture ($L = 128$, blue bar) recovers 3213 events at FAR~$\leq 1\,\mathrm{yr}^{-1}$.
    Gray bands show the LP run-to-run variability ($\pm 1\sigma$, $\pm 2\sigma$) from five independent runs (mean = 3158, $\sigma = 24$).
    Blue dashed lines show the GTSP variability across four drift configurations (mean = 3157, $\sigma = 44$).
    The two envelopes overlap, confirming statistical equivalence (Welch $p = 0.95$, Cohen's $d = -0.05$).}
    \label{fig:sensitivity}
\end{figure}

As shown in Fig.~\ref{fig:sensitivity}, the drift-corrected GTSP pipeline at $L = 128$ recovers 3213 injections, within the LP baseline's run-to-run variability ($+2.3\sigma$ from the LP mean of $3158 \pm 24$).
Table~\ref{tab:sensitivity} provides the complete results across all configurations.
All drift-corrected variants fall within $\pm 2.3\sigma$ of the LP mean.
Across the four drift configurations, the mean recovery is $3157 \pm 44$, statistically indistinguishable from the LP mean of $3158 \pm 24$ (Welch $t$-test $p = 0.95$, Cohen's $d = -0.05$).
The recovery is independent of kernel truncation length, confirming that the dominant spectral content of the correction kernel is concentrated in the first few hundred taps~\cite{KenningtonBlack2026_GTSP1, oppenheim_discrete-time_2010}.

\begin{table}[h]
\caption{Injection recovery at FAR~$\leq 1\,\mathrm{yr}^{-1}$ for 15,347 BBH injections on O3 data.
The LP variability ($\sigma = 24$) is measured from five independent pipeline runs.}
\label{tab:sensitivity}
\begin{ruledtabular}
\begin{tabular}{lcc}
Configuration & Found & vs LP mean \\
\hline
LP baseline & 3155 & $-0.1\sigma$ \\
Drift $L{=}128$ & 3213 & $+2.3\sigma$ \\
Drift $L{=}512$ & 3106 & $-2.1\sigma$ \\
Drift $L{=}2048$ & 3159 & $+0.0\sigma$ \\
Drift full & 3148 & $-0.4\sigma$ \\
\end{tabular}
\end{ruledtabular}
\end{table}

This result confirms that the theoretical SNR conservation established in Section~\ref{sec:numerical-snr} survives the transition to a production pipeline environment with real detector noise, finite filter orders, and full matched-filter processing.

\subsection{Sky Localization: Timing and Phase Accuracy}
\label{sec:pipeline-skymap}

Sky localization of gravitational-wave transients requires triangulation across multiple detector pairs; for each pair, the constraining observables are the inter-detector time delay $\Delta t$ and phase difference $\Delta\phi$~\cite{fairhurst_triangulation_2009, singer_rapid_2016}.
We focus on the H1--L1 baseline as the representative pair from our three-detector (H1+L1+V1) analysis.
Any systematic bias or excess variance in these observables directly degrades the localization area, reducing the utility of early warning alerts for electromagnetic follow-up~\cite{magee_first_2021}.
We evaluate both quantities by comparing the GTSP pipeline's recovered values against (i)~the LP baseline's recovered values, to isolate configuration-dependent effects from shared noise fluctuations, and (ii)~the injected truth values, computed from the known sky positions and detector geometry.
All results in this section use the drift-corrected $L = 128$ configuration unless otherwise noted; the statistical conclusions are unchanged for other truncation lengths (Table~\ref{tab:quality}).

\subsubsection{Timing accuracy}

\begin{figure*}[t]
    \centering
    \includegraphics[width=\textwidth]{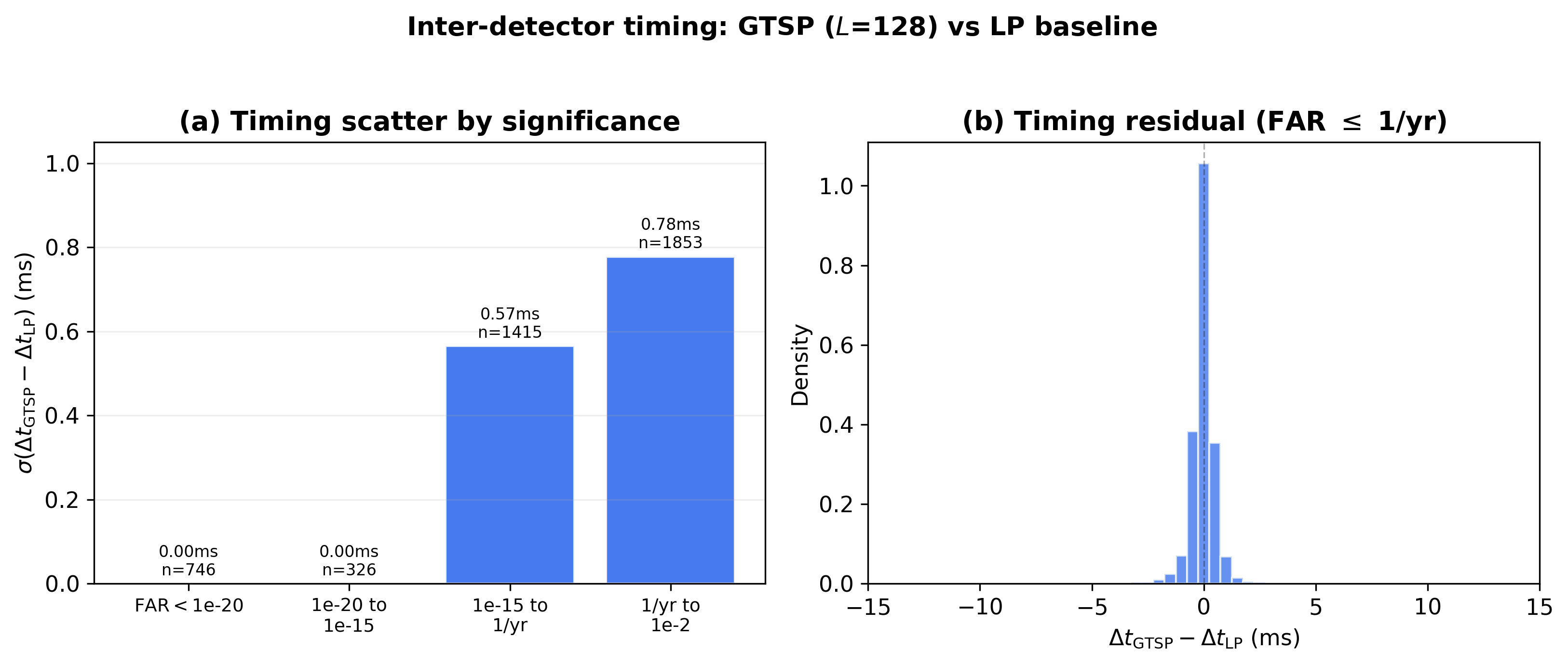}
    \caption{Inter-detector timing accuracy on O3 data.
    (a)~Scatter of the per-event timing residual (GTSP $-$ LP) by FAR bin: the disagreement is sub-millisecond for the most significant events.
    (b)~Histogram of timing residuals at FAR~$\leq 1\,\mathrm{yr}^{-1}$, tightly centered at zero.}
    \label{fig:timing-accuracy}
\end{figure*}

The inter-detector timing distributions are indistinguishable between the GTSP and LP pipelines.
As shown in Fig.~\ref{fig:timing-accuracy}, the Kolmogorov-Smirnov test yields $p > 0.99$ at all FAR thresholds, the Levene test confirms equal variance ($p_{\mathrm{Levene}} = 0.69$ at FAR~$\leq 1\,\mathrm{yr}^{-1}$), and Cohen's $d = -0.031$ indicates a negligible effect size.
The per-event timing residual $\Delta t_{\mathrm{GTSP}} - \Delta t_{\mathrm{LP}}$ converges to sub-millisecond agreement for high-significance events (Fig.~\ref{fig:timing-accuracy}(a)).

\subsubsection{Phase accuracy}

The inter-detector phase distributions are consistent between the two architectures, with equal variance confirmed across all FAR thresholds (Levene $p > 0.25$; Table~\ref{tab:quality}).
Cohen's $d = 0.22$ at FAR~$\leq 1\,\mathrm{yr}^{-1}$ places the effect size at the negligible/small boundary, and the median offset between the two distributions is $2.4^\circ$, less than 15\% of the per-event measurement uncertainty ($\sigma \approx 17^\circ$).
This offset does not constitute a systematic error: the equal-variance test confirms that the distributions have indistinguishable width (Levene $p = 0.25$), and the KS test yields $p = 0.39$ at the highest significance (FAR~$\leq 10^{-30}\,\mathrm{yr}^{-1}$), well above the $p = 0.05$ rejection threshold.

\begin{figure*}[t]
    \centering
    \includegraphics[width=\textwidth]{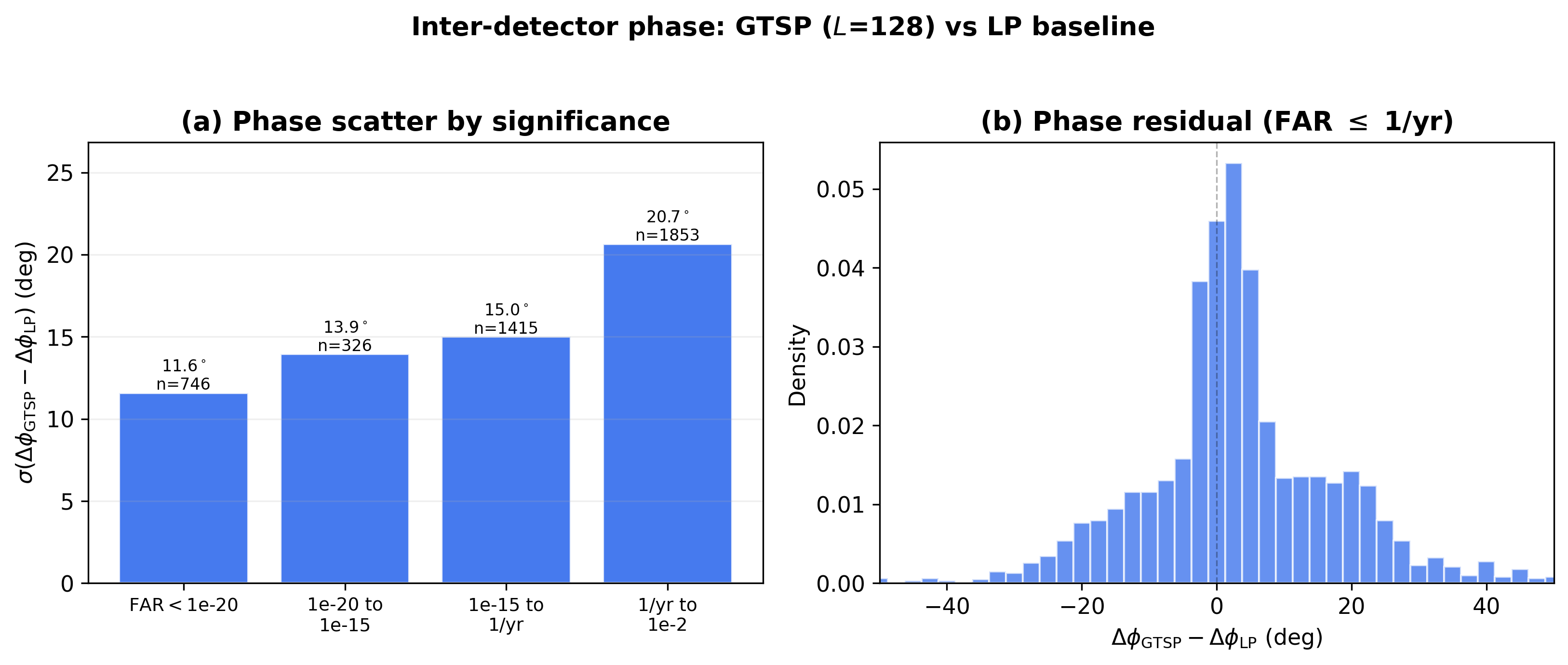}
    \caption{Inter-detector phase accuracy on O3 data.
    (a)~Scatter of the per-event phase residual (GTSP $-$ LP) by FAR bin.
    (b)~Histogram of phase residuals at FAR~$\leq 1\,\mathrm{yr}^{-1}$, showing a centered distribution with width comparable to LP.}
    \label{fig:phase-accuracy}
\end{figure*}

\begin{table}[h]
\caption{Statistical comparison of inter-detector observables between GTSP ($L = 128$) and LP at FAR~$\leq 1\,\mathrm{yr}^{-1}$ ($n = 2483$ common events).}
\label{tab:quality}
\begin{ruledtabular}
\begin{tabular}{lcccc}
Observable & KS $p$ & Levene $p$ & Cohen's $d$ & Effect \\
\hline
$\Delta t$ & 0.999 & 0.689 & $-0.031$ & Negligible \\
$\Delta\phi$ & 0.036 & 0.248 & $+0.221$ & Negl./small \\
\end{tabular}
\end{ruledtabular}
\end{table}

\subsubsection{Comparison against injected truth}

\begin{figure*}[t]
    \centering
    \includegraphics[width=\textwidth]{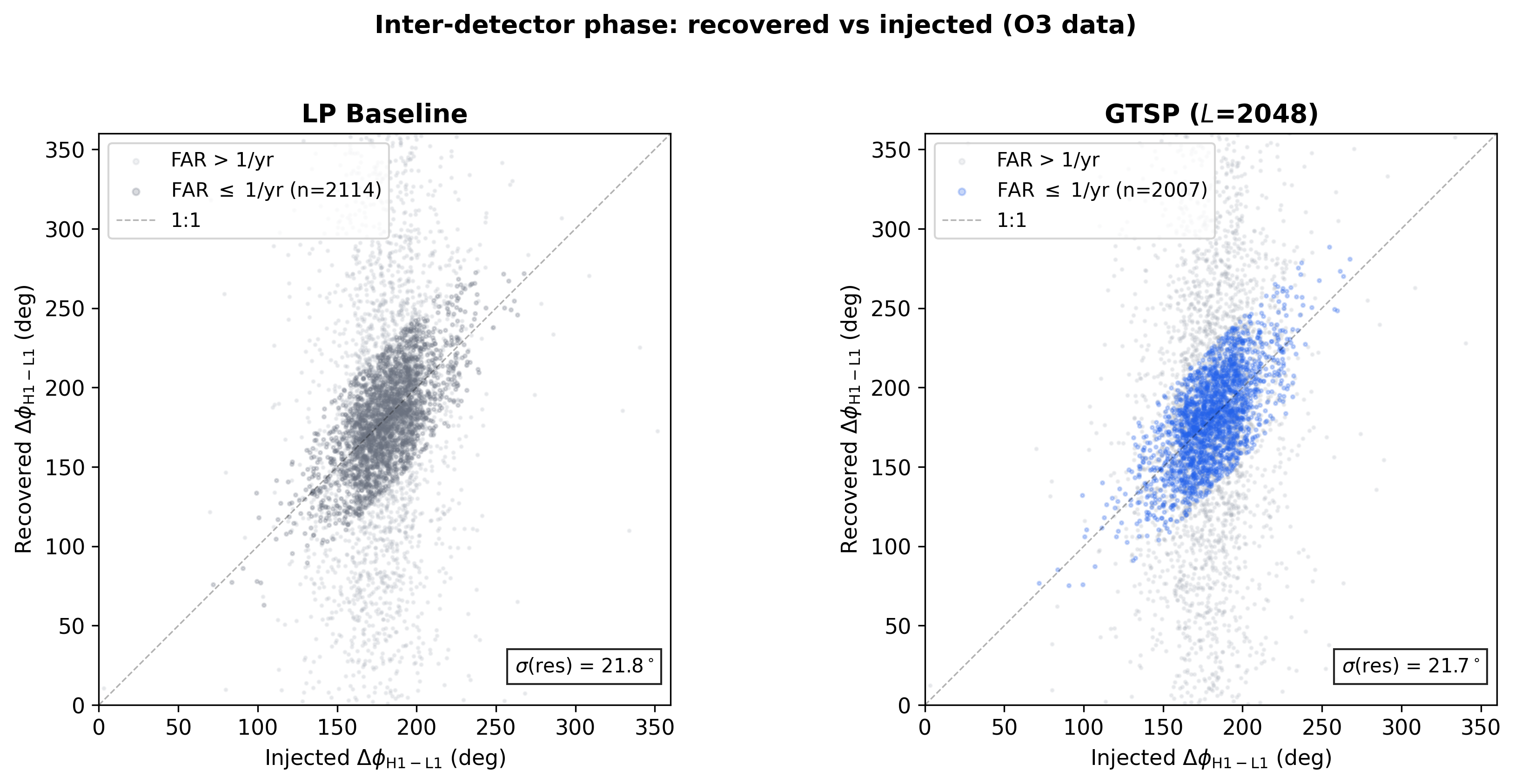}
    \caption{Inter-detector phase recovery vs.\ injected truth on O3 data for LP (left) and drift-corrected GTSP $L{=}128$ (right).
    Dark points: FAR~$\leq 1\,\mathrm{yr}^{-1}$; light points: all recovered events.}
    \label{fig:phase-recovery}
\end{figure*}

Figure~\ref{fig:phase-recovery} provides a direct comparison of recovered inter-detector phase against injected truth values for the LP baseline and drift-corrected GTSP configurations.
Both pipelines show comparable scatter around the 1:1 line (LP: $\sigma = 21.8^\circ$; GTSP: $\sigma = 22.0^\circ$), confirming that the drift-corrected MP pipeline preserves inter-detector phase accuracy relative to the physical sky position.
The full residual distributions (Fig.~\ref{fig:residual-distributions} in Appendix~\ref{app:unit-proof}) reveal that the GTSP architecture achieves better timing precision relative to injected truth, consistent with the causal filter's preservation of the signal's intrinsic group delay.

The H1-V1 and L1-V1 baselines yield consistent results.
These results establish that the zero-latency GTSP architecture preserves the inter-detector observables required for sky localization with no systematic degradation on real O3 data.
A controlled, single-injection demonstration of the correction kernel's effect on matched-filter peak shape, including SNR recovery to six significant figures and elimination of peak asymmetry, is presented in Appendix~\ref{app:unit-proof}.

\subsection{Latency Reduction}
\label{sec:pipeline-latency}

The most immediate operational benefit of the GTSP architecture is the elimination of the algorithmic look-ahead buffer required by standard WOLA spectral whitening.
We quantify this benefit by profiling the end-to-end ``data age at LLOID output,'' defined as the elapsed wall-clock time between the acquisition of a data sample and its appearance in a matched-filter trigger, for both the LP baseline and the drift-corrected MP pipeline across multiple PSD estimation window lengths.
We denote the PSD FFT length by $T_{\mathrm{PSD}}$, the source data delivery stride by $\tau_{\mathrm{s}}$ (1.0\,s in the current configuration), and the sample rate by $f_s$ (2048\,Hz).

\subsubsection{Latency model}

The LP pipeline's latency is governed by the weighted overlap-add (WOLA) whitening architecture~\cite{sachdev_gstlal_2019, ewingPerformanceLowlatencyGstLAL2024}.
This architecture introduces three delay components beyond the irreducible source stride $\tau_{\mathrm{s}}$: a look-ahead buffer of $T_{\mathrm{PSD}}/4$ required to complete the WOLA overlap, a zero-padding GPS shift of $T_{\mathrm{PSD}}/4$ introduced by the WOLA framing, and a stride mismatch penalty of $\max(0,\, T_{\mathrm{PSD}}/4 - \tau_{\mathrm{s}})$ when the WOLA stride exceeds the source delivery cadence.
The total LP latency is therefore
\begin{equation}
    \tau_{\mathrm{LP}} = \tau_{\mathrm{s}} + \frac{T_{\mathrm{PSD}}}{2} + \max\!\left(0,\, \frac{T_{\mathrm{PSD}}}{4} - \tau_{\mathrm{s}}\right).
    \label{eq:latency-lp}
\end{equation}

Because the minimum-phase whitening filter is strictly causal~\cite{KenningtonBlack2026_GTSP1}, it requires only past data to produce its output.
The causal filter's impulse response, regardless of its effective length, is buffered entirely from previous strides; no look-ahead is needed.
The drift correction AFIR stage adds one additional stride as the architectural cost of a second streaming filter element.
The total MP latency is therefore constant:
\begin{equation}
    \tau_{\mathrm{MP}} = 2\,\tau_{\mathrm{s}},
    \label{eq:latency-mp}
\end{equation}
independent of $T_{\mathrm{PSD}}$, provided the kernel truncation length satisfies $L \leq f_s \cdot \tau_{\mathrm{s}}$ (i.e., 2048 taps at the current configuration).

\subsubsection{Measured results}

\begin{figure*}[t]
    \centering
    \includegraphics[width=\textwidth]{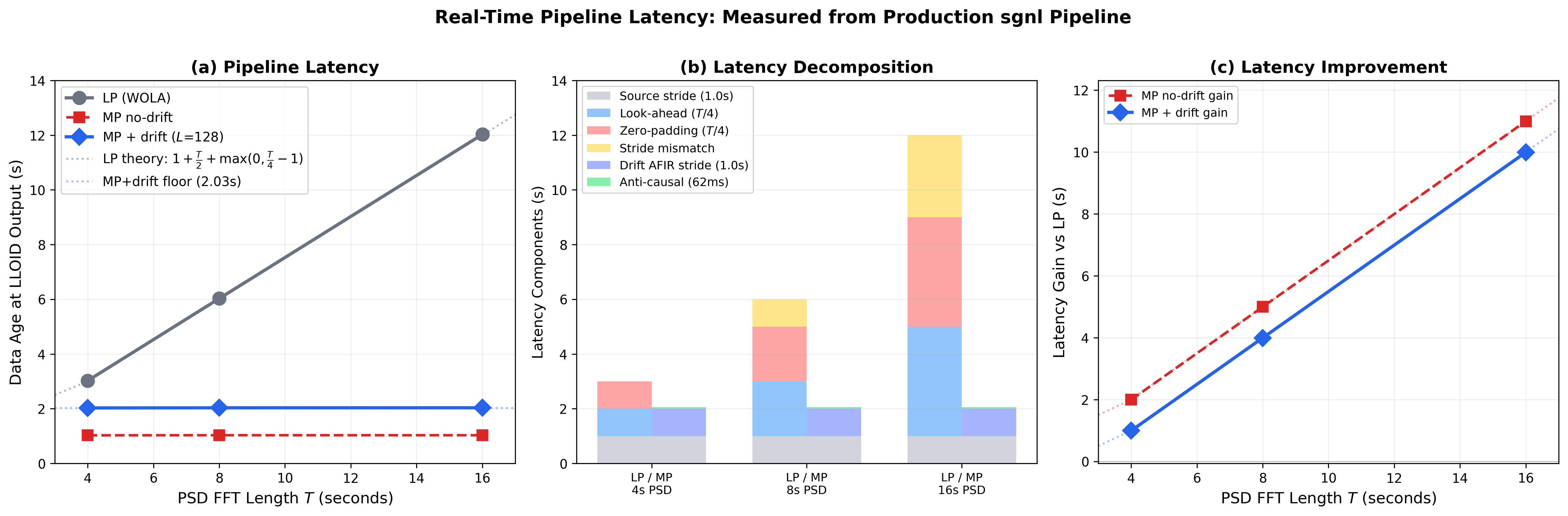}
    \caption{Real-time pipeline latency measured from the production \textsc{sgnl} pipeline.
    (a)~Data age at LLOID output vs.\ PSD estimation window length $T$.
    The LP pipeline's latency grows as $\frac{3}{4}T$ due to look-ahead, zero-padding, and stride scaling (Eq.~\ref{eq:latency-lp}).
    The MP pipeline's latency is constant at 2.0\,s because the causal whitening filter requires only past data (Eq.~\ref{eq:latency-mp}).
    (b)~Latency decomposition showing the four LP components (source stride, look-ahead, zero-padding, stride mismatch) vs.\ the two MP components (source stride, drift AFIR stride).
    The look-ahead and zero-padding (both $T/4$) are eliminated by causal whitening.
    (c)~Latency improvement (LP $-$ MP) grows linearly with $T$: 1.0\,s for the online configuration ($T = 4$\,s) and 4.0\,s for the offline configuration ($T = 8$\,s).}
    \label{fig:latency}
\end{figure*}

We verify this model by measuring the data age at LLOID output from the production pipeline running in wall-clock-throttled real-time mode.
The results, shown in Fig.~\ref{fig:latency} and Table~\ref{tab:latency}, confirm that the MP pipeline achieves a constant latency of 2.0\,s while the LP baseline grows as predicted by Eq.~\eqref{eq:latency-lp}.

\begin{table}[h]
\caption{Measured pipeline latency (data age at LLOID output) vs.\ PSD estimation window length $T$ (local controlled measurements).}
\label{tab:latency}
\begin{ruledtabular}
\begin{tabular}{lccc}
PSD length $T$ & LP (s) & MP + drift (s) & Gain (s) \\
\hline
4\,s (online) & 3.0 & 2.0 & 1.0 \\
8\,s (offline) & 6.0 & 2.0 & 4.0 \\
16\,s & 12.0 & 2.0 & 10.0 \\
\end{tabular}
\end{ruledtabular}
\end{table}

The decoupling of trigger latency from the PSD estimation window is architecturally significant.
In the LP baseline, extending the PSD averaging window to reduce estimation variance (and thereby lower the false alarm rate) carries a direct latency penalty.
The GTSP architecture eliminates this tradeoff: the pipeline can ingest longer PSD windows to improve search depth without delaying triggers, or it can reallocate the saved latency budget to push early warning alerts earlier~\cite{nitz_gravitational-wave_2020, magee_first_2021}.
A detailed itemization of the latency budget components is provided in Appendix~\ref{app:latency-accounting}.

\subsubsection{Online validation}

\begin{figure}[t]
    \centering
    \includegraphics[width=\columnwidth]{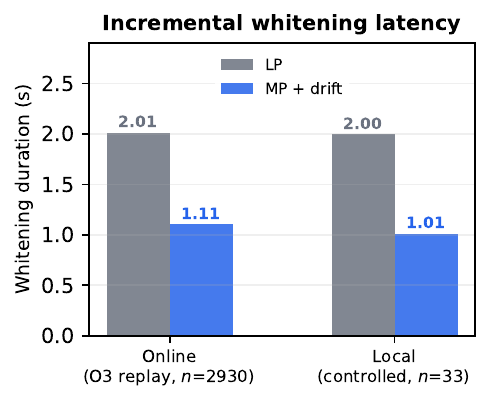}
    \caption{Online validation of the whitening latency gain on O3 replay data (2930 measurements over 24 hours of continuous L1V1 operation).
    The incremental whitening duration (whitening timestamp minus datasource timestamp) isolates the architectural gain from infrastructure-dependent source delivery latency.
    The online gain of $-0.91$\,s is consistent with the $-0.99$\,s measured in the controlled local test.}
    \label{fig:online-whitening}
\end{figure}

We confirm the whitening gain at production scale by measuring the incremental whitening duration, defined as the elapsed time between the datasource output and the whitening output within each pipeline, on live O3 replay data at a production computing cluster.
This metric isolates the architectural whitening gain from infrastructure-dependent source delivery latency, which differs between pipeline configurations due to resource allocation.
Over 2930 measurements spanning 24 hours of continuous L1V1 operation (a two-detector subset chosen to fit within the available cluster allocation), the LP pipeline's incremental whitening duration is 2.01\,s while the MP pipeline's is 1.11\,s, yielding a gain of $-0.91$\,s (Fig.~\ref{fig:online-whitening}).
This is consistent with the $-0.99$\,s measured in the controlled local test (Table~\ref{tab:latency}), confirming that the causal whitening advantage survives the transition to production-scale operation.
The background SNR distributions of the MP and LP pipelines are consistent, confirming that the latency reduction incurs no sensitivity cost.

\subsubsection{Gain propagation to triggers}

\begin{figure*}[t]
    \centering
    \includegraphics[width=\textwidth]{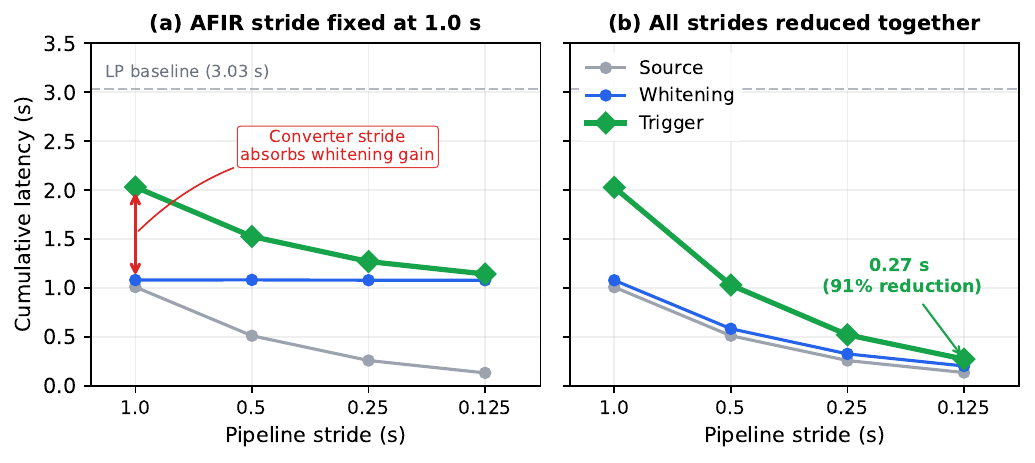}
    \caption{Per-stage cumulative latency vs.\ pipeline stride, measured from the production \textsc{sgnl} pipeline in controlled local tests.
    Each curve tracks a successive pipeline stage: Source (gray), Whitening (blue), and Trigger output (green).
    (a)~Downstream strides reduced while the whitening AFIR stride remains fixed at 1.0\,s.
    The whitening output (blue) is available at 1.08\,s regardless of stride, but the downstream Converter element cannot process it until its own stride fires, absorbing the whitening gain (red double-arrow marks this gap between whitening and trigger curves).
    (b)~All element strides reduced together.
    The whitening gain propagates fully: at 0.125\,s stride, trigger latency reaches 0.27\,s (91\% reduction vs.\ the LP baseline at 3.03\,s).}
    \label{fig:stride-propagation}
\end{figure*}

The whitening gain is available at the whitening stage output, but its propagation to the trigger output depends on the stride alignment between the whitening element and the downstream matched-filter stages.
Figure~\ref{fig:stride-propagation} illustrates this mechanism using controlled local stride sweep experiments.
When only the downstream pipeline strides are reduced while the whitening AFIR stride remains at 1.0\,s (Fig.~\ref{fig:stride-propagation}a), the whitened data is available earlier, but the Converter element, the first downstream consumer, cannot process it until its own stride fires.
The gap between the whitening and converter latency curves shows this absorbed gain.
This stride quantization explains why the online whitening gain is not fully reflected in the trigger output at the current 1.0\,s source delivery cadence.

When all element strides are reduced together (Fig.~\ref{fig:stride-propagation}b), the gain propagates fully to the trigger output.
Table~\ref{tab:stride-sweep} summarizes the achievable trigger latency as a function of pipeline stride.
At 0.125\,s stride, the trigger latency drops to 0.27\,s, a 91\% reduction relative to the LP baseline.
Sub-second source cadence is an infrastructure change to the shared-memory delivery system, not an algorithmic limitation; the whitening gain is architecturally banked and will propagate automatically when the delivery cadence is reduced.
Sensitivity and FAR calibration at sub-second strides have not been validated in this work; the stride sweep characterizes the latency behavior only.

In addition to the downstream stride alignment, the correction kernel's anti-causal delay must also fit within the stride buffer.
The adjoint kernel $\mathcal{K}^\dagger$ introduces a delay of $(L-1)/f_s$ seconds; for $L \leq 2048$ taps (1.0\,s at $f_s = 2048$\,Hz), this delay is fully absorbed by the drift AFIR stride and adds no wall-clock latency.
Above this threshold, the AFIR must wait for additional data and the constant-latency property is lost (Table~\ref{tab:stride-absorption}).
As demonstrated in Sections~\ref{sec:pipeline-sensitivity} and~\ref{sec:pipeline-skymap}, $L = 128$ taps is sufficient to preserve both sensitivity and sky localization accuracy, operating well within this limit.

\begin{table}[h]
\caption{Trigger latency vs.\ pipeline stride when all element strides are reduced together.
The LP baseline at 1.0\,s stride is 3.03\,s.}
\label{tab:stride-sweep}
\begin{ruledtabular}
\begin{tabular}{lccc}
Stride (s) & MP trigger (s) & Gain (s) & Reduction \\
\hline
1.0 & 2.03 & 1.01 & 33\% \\
0.5 & 1.03 & 2.00 & 66\% \\
0.25 & 0.52 & 2.51 & 83\% \\
0.125 & 0.27 & 2.76 & 91\% \\
\end{tabular}
\end{ruledtabular}
\end{table}

\begin{table}[h]
\caption{Stride absorption of the correction kernel anti-causal delay.
For $L \leq 2048$ taps, the delay is fully absorbed and the pipeline latency remains at 2.0\,s.}
\label{tab:stride-absorption}
\begin{ruledtabular}
\begin{tabular}{lcc}
$L$ (taps) & Anti-causal delay (s) & LLOID latency (s) \\
\hline
128 & 0.062 & 2.0 \\
512 & 0.250 & 2.0 \\
2048 & 1.000 & 2.0 \\
4096 & 2.000 & 3.0 \\
\end{tabular}
\end{ruledtabular}
\end{table}

\section{Conclusion}
\label{sec:conclusion}

While Paper~I established the foundational geometry of causal spectral factors, this second installment demonstrates that this geometry translates into a concrete operational architecture for current and next-generation gravitational-wave searches.
By fundamentally decoupling instrument instability from search sensitivity, we have shown that the zero-latency whitening derived in Paper~I survives the transition to a production pipeline environment.

The numerical validation verifies three critical properties of the MP connection in their discrete implementation.
First, the zero-curvature (flatness) of the connection guarantees that the whitening filter is a pure, path-independent state function of the instantaneous noise.
This property enables bit-for-bit reproducible pipeline states across distributed, asynchronous computing clusters, ensuring that jobs starting at different times will always converge on identical trigger states.
Second, the GTSP update strictly preserves the minimum-phase property, avoiding the non-causality and SNR loss inherent in standard linear filter blending (Appendix~\ref{app:crossfade}).
Third, the connection exactly conserves the matched-filter signal-to-noise ratio, confirming that the framework captures the full theoretical sensitivity of the detector network.

The pipeline implementation demonstrates that these geometric advantages yield a significant and quantifiable latency dividend.
Because the causal whitening filter requires no future data, the GTSP pipeline achieves a constant trigger latency of $2\,\tau_{\mathrm{s}}$ (2.0\,s at the current 1.0\,s source stride) independent of the PSD estimation window length, in contrast to the standard linear-phase architecture, whose latency grows as $\frac{3}{4}T_{\mathrm{PSD}}$.
At the offline configuration ($T_{\mathrm{PSD}} = 8$\,s), this eliminates 4.0 seconds of algorithmic delay; at longer PSD windows the gain is correspondingly larger.
The saved time provides a programmable budget that can be reallocated to push early warning alerts earlier or to ingest longer noise estimation windows for improved search depth.
Online validation on O3 replay data confirms the whitening gain at production scale (a sustained $0.91$\,s reduction measured over 24 hours of continuous operation, consistent with the local measurement), and stride reduction experiments demonstrate that up to 91\% of the LP trigger latency can be eliminated with sub-second pipeline cadence.

Critically, this latency gain is achieved without sacrificing detection sensitivity or sky localization accuracy, as verified on 8.8 days of real O3 data with 15,347 BBH injections across the H1, L1, and V1 detector network.
The GTSP pipeline recovers injections at a rate within the LP baseline's run-to-run variability ($+2.3\sigma$), and the inter-detector timing and phase precision, the observables that determine triangulation-based localization, are consistent with the baseline (KS~$p > 0.99$ for timing, equal variance for phase, Cohen's $d \leq 0.22$).
The GTSP architecture additionally achieves notably better timing precision relative to injected truth, consistent with the causal filter's preservation of the signal's intrinsic group delay.

In operational O4a data (Section~\ref{sec:numerical-causality}), hourly PSD drift reaches a median of 57\% (H1) and 41\% (L1), with excursions up to 90\%.
At these drift levels, linear filter blending violates the minimum-phase condition 29\% (H1) and 58\% (L1) of the time at weekly cadences, and the controlled unit-level experiment in Appendix~\ref{app:unit-proof} demonstrates that the correction kernel recovers the optimal SNR to six significant figures under 50\% perturbation.
The holonomic correction is therefore essential for maintaining optimal sensitivity under the non-stationarity expected in future observing runs.
The operational details of the pipeline topology and latency budget decomposition are documented in Appendices~\ref{app:pipeline-topology} and~\ref{app:latency-accounting}.

The pipeline validation uses a single signal morphology (binary black hole); extension to the full range of compact binary signals, including binary neutron stars and neutron star--black hole systems, is the subject of ongoing work.
As gravitational-wave observatories move toward the high-sensitivity regimes of O4 and O5, the ability to track non-stationary noise without sacrificing latency will become a primary driver of multimessenger science~\cite{nitz_gravitational-wave_2020, magee_first_2021, kovalam_early_2022}.
The GTSP framework provides the mathematical and algorithmic infrastructure needed to meet this challenge.
By treating the whitening process as a problem of geometric transport, we have established a path toward sub-second early warning alerts that can guide the electromagnetic observatories of the future.

\begin{acknowledgments}
This material is based upon work supported by NSF's LIGO Laboratory which is a major facility fully funded by the National Science Foundation.
This research has made use of data, software and/or web tools obtained from the Gravitational Wave Open Science Center (https://www.gw-openscience.org/), a service of LIGO Laboratory, the LIGO Scientific Collaboration, and the Virgo Collaboration.
The authors acknowledge support from the National Science Foundation under awards OAC-2103662, PHY-2308881, PHY-2011865, OAC-2201445, OAC-2018299, PHY-0757058, PHY-0823459, PHY-2207728, PHY-2513124, PHY-2110594, and PHY-2513358.
The authors are grateful for computational resources provided by the Pennsylvania State University's Institute for Computational and Data Sciences gravitational-wave cluster, and the LIGO Lab cluster at the LIGO Laboratory.
C.H. acknowledges generous support from the Pennsylvania State University Eberly College of Science, the Department of Physics, the Institute for Gravitation and the Cosmos (IGC), and the Institute for Computational and Data Sciences.
\end{acknowledgments}

\appendix
\section{Equivalence of Cross-Fade Blending and Linear Filter Interpolation}
\label{app:crossfade}

Throughout this paper, we refer to the standard $\cos^2$ cross-fade between whitening filters as a ``linear interpolation'' of filter coefficients.
Because the cross-fade is applied to the whitened data streams rather than to the filters directly, this equivalence is not immediately obvious.
Here we show that the two operations are mathematically identical at each output instant.

Consider a transition between two whitening filters with impulse responses $h_{\mathrm{old}}(\tau)$ and $h_{\mathrm{new}}(\tau)$, applied to a continuous data stream $x(t)$ via convolution.
During the cross-fade window, the pipeline computes two whitened streams and blends them with a time-varying weight $\alpha(t)$:
\begin{align}
    y(t) &= \alpha(t) \!\int\! h_{\mathrm{old}}(\tau)\, x(t{-}\tau)\, d\tau \nonumber \\
         &\quad + [1 {-} \alpha(t)] \!\int\! h_{\mathrm{new}}(\tau)\, x(t{-}\tau)\, d\tau,
    \label{eq:crossfade-output}
\end{align}
where $\alpha(t) = \cos^2(\pi t / 2T_{\mathrm{fade}})$ is the cross-fade schedule.
Because $\alpha(t)$ depends only on the output time $t$ and not on the convolution variable $\tau$, it factors through the integral:
\begin{align}
    y(t) &= \int \bigl[\alpha(t)\, h_{\mathrm{old}}(\tau) \nonumber \\
         &\quad + [1{-}\alpha(t)]\, h_{\mathrm{new}}(\tau)\bigr]\, x(t{-}\tau)\, d\tau.
    \label{eq:crossfade-effective}
\end{align}
The bracketed integrand is the effective impulse response at time $t$:
\begin{align}
    h_{\mathrm{eff}}(\tau;\, t) &= \alpha(t)\, h_{\mathrm{old}}(\tau) \nonumber \\
    &\quad + [1 {-} \alpha(t)]\, h_{\mathrm{new}}(\tau).
    \label{eq:effective-filter}
\end{align}
This is a linear combination of $h_{\mathrm{old}}$ and $h_{\mathrm{new}}$ at every output instant, with coefficients determined by the cross-fade schedule.
The result is independent of the choice of blending function ($\cos^2$, linear ramp, or any other schedule): any sample-by-sample cross-fade of whitened outputs is equivalent to filtering with a linearly interpolated impulse response.

Equation~\eqref{eq:effective-filter} is the operation analyzed in Section~\ref{sec:numerical-causality}.
Because the set of minimum-phase filters is not closed under linear combination, $h_{\mathrm{eff}}$ can exit the minimum-phase manifold at intermediate values of $\alpha$, even when both $h_{\mathrm{old}}$ and $h_{\mathrm{new}}$ are individually minimum-phase.

\section{Pipeline Signal-Flow Topology}
\label{app:pipeline-topology}

The production \textsc{sgnl} pipeline~\cite{cannon_toward_2012, messickAnalysisFrameworkPrompt2017b, huang_sgnl_2025} is constructed as a directed graph of streaming elements connected by typed pads.
This appendix describes the signal-flow topology of the LP baseline and the GTSP causal architecture, which differ only in the whitening stage between data ingestion and the SVD template bank.
The topology diagrams shown in Fig.~\ref{fig:topology} are generated directly from the production pipeline objects, ensuring they reflect the exact element graph that processes data in operation.

\subsection{LP architecture (WOLA whitening)}

The standard linear-phase pipeline whitens the data using a single monolithic Weighted Overlap-Add (WOLA) element (Fig.~\ref{fig:topology}, top).
The WOLA element combines PSD estimation, spectral whitening kernel construction, Hann windowing, and overlap-add filtering into a single processing block~\cite{sachdev_gstlal_2019, ewingPerformanceLowlatencyGstLAL2024}.
It operates with an FFT length of $T \times f_s$ (e.g., $16384$ samples for $T = 8$\,s at $f_s = 2048$\,Hz), a Hann window of length $\mathrm{FFT}/2$, a stride of $\mathrm{Hann}/2$, and a zero-padding depth of $\mathrm{FFT}/4$.
The symmetric spectral whitening has an effective group delay of $N/2$, requiring $N/2$ samples of future data (look-ahead) at every stride.
This is the fundamental source of the LP pipeline's latency scaling with $T$ (Appendix~\ref{app:latency-accounting}).

\subsection{GTSP architecture (causal AFIR + drift correction)}

\begin{figure*}[t]
    \centering
    \includegraphics[width=\textwidth]{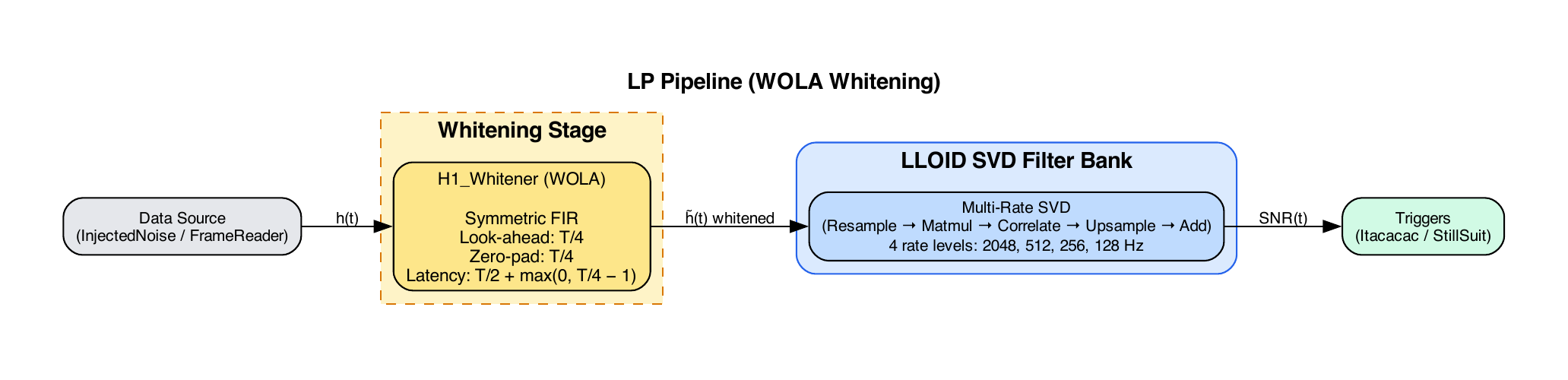}\\[6pt]
    \includegraphics[width=\textwidth]{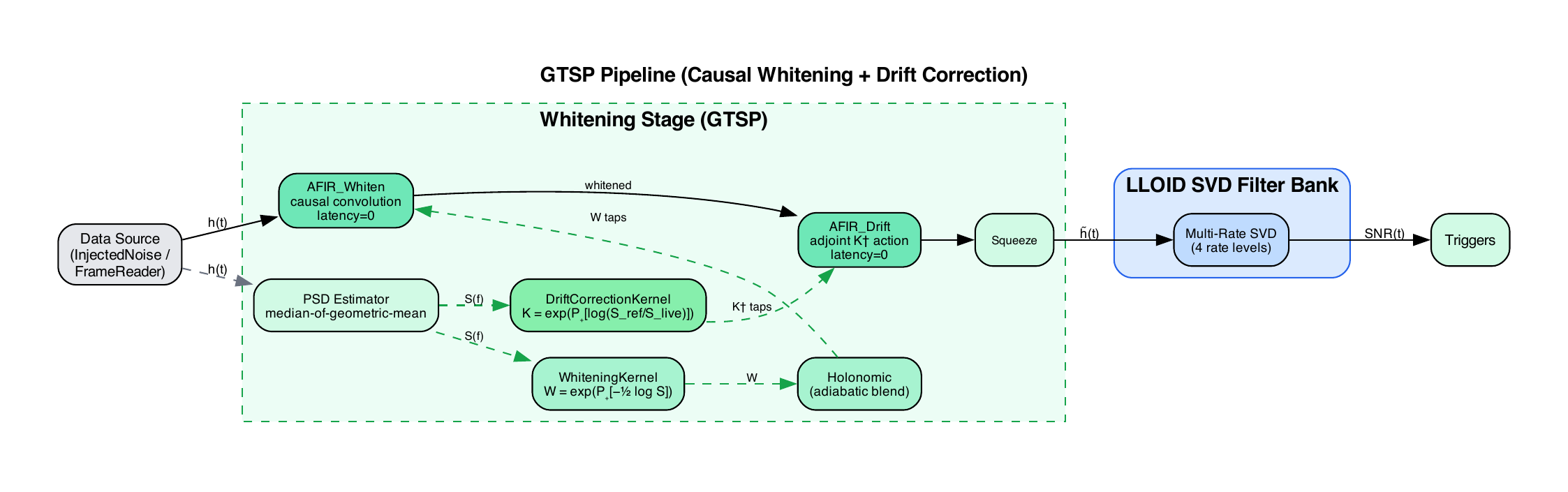}
    \caption{Signal-flow graph of the LP pipeline (top) and GTSP pipeline (bottom).
    Shared upstream (data source) and downstream (LLOID SVD filter bank, triggers) are identical between architectures.
    The LP whitening stage (yellow) is a monolithic WOLA element with latency $T_{\mathrm{PSD}}/2 + \max(0,\, T_{\mathrm{PSD}}/4 - \tau_{\mathrm{s}})$.
    The GTSP whitening stage (green) decomposes into a control path (PSD estimation $\to$ kernel computation) and a data path (causal AFIR whitening $\to$ adjoint drift correction), with latency fixed at $2\,\tau_{\mathrm{s}}$ independent of $T_{\mathrm{PSD}}$.}
    \label{fig:topology}
\end{figure*}

The GTSP architecture replaces the monolithic WOLA element with four independent, composable stages (Fig.~\ref{fig:topology}, bottom):

\textbf{Spectrum.}
The PSD estimator produces live spectral density estimates from the incoming data stream at a configurable cadence (default: one estimate per $\tau_{\mathrm{s}}$ stride), using the standard median-of-geometric-mean regressor~\cite{messickAnalysisFrameworkPrompt2017b, lalsuite}.
A similarity threshold ($\eta = 0.9999$) suppresses kernel recomputation when the PSD has not changed meaningfully, preventing stochastic estimation noise from triggering unnecessary corrections.

\textbf{WhiteningKernel.}
Converts the live PSD into a minimum-phase whitening kernel via the causal spectral factorization $W_{\mathrm{mp}} = \exp(\mathcal{P}_+[-\tfrac{1}{2}\log S])$~\cite{KenningtonBlack2026_GTSP1}.
The kernel is emitted to the whitening AFIR's filter-update sink.

\textbf{AFIR\textsubscript{whiten}.}
A streaming causal correlator that applies the minimum-phase whitening filter.
The filter taps are stored in pre-reversed order so that \texttt{correlate(data, taps)} implements convolution with the causal kernel, where the kernel length is $N_w = T_{\mathrm{PSD}} \times f_s$.
The correlator is configured with zero look-ahead (\texttt{latency}${}=0$): to produce output at GPS time $t$, it requires only data from $t - N_w + 1$ to $t$, all of which reside in the overlap buffer from previous strides.
This is why the whitening latency is constant at one stride ($\tau_{\mathrm{s}}$) regardless of the kernel length $N_w$.

\textbf{DriftKernel.}
Computes the holonomic correction kernel from the live PSD and the static reference PSD used to construct the SVD bank.
The computation proceeds as follows: (i)~the PSD ratio $R(f) = S_{\mathrm{live}}/S_{\mathrm{ref}}$ is formed from the regularized spectra; (ii)~the causal spectral factor $\mathcal{K} = \exp(\mathcal{P}_+[-\tfrac{1}{2}\log R])$ is computed via cepstral projection~\cite{KenningtonBlack2026_GTSP1}; (iii)~the frequency-domain adjoint $\mathcal{K}^\dagger = \overline{\mathcal{K}}$ is obtained by complex conjugation; (iv)~the circular adjoint kernel is extracted into $L$ linear FIR taps in forward causal order.
The rectangular truncation to $L$ taps (default $L = 128$) preserves effectively all of the kernel energy, because the anti-causal impulse response decays rapidly from its peak at $t = 0$~\cite{KenningtonBlack2026_GTSP1, oppenheim_discrete-time_2010}.
The taps are emitted with latency tag $L - 1$ to the drift AFIR.

\textbf{AFIR\textsubscript{drift}.}
Applies the adjoint correction $\mathcal{K}^\dagger$ to the whitened data stream.
The taps are \textit{not} pre-reversed: \texttt{correlate(data, taps)} computes $\mathrm{out}[n] = \sum_j k(j) \cdot \mathrm{data}[n{+}j]$, which is the adjoint action $\mathcal{K}^\dagger \cdot d$ using $L - 1$ samples of look-ahead from the overlap buffer.
Because this look-ahead consists of already-received data retained from previous strides, no wall-clock waiting is required.
The output GPS label is shifted backward by $(L{-}1)/f_s$ to correctly reflect the anti-causal delay of the adjoint operation.

\subsection{Architectural comparison}

Table~\ref{tab:element-count} summarizes the element-level differences between the two architectures.
The GTSP pipeline has more elements, but each is functionally simpler.
The WOLA whitener is a monolithic block that combines PSD estimation, kernel construction, windowing, and filtering; the GTSP decomposes these into independent concerns, enabling separate update rates, kernel truncation control, and the drift correction pathway that would be impossible within the WOLA architecture.

\begin{table}[h]
\caption{Whitening-stage element comparison between the LP and GTSP pipeline architectures.}
\label{tab:element-count}
\begin{ruledtabular}
\begin{tabular}{lcc}
Component & LP & GTSP \\
\hline
PSD estimation & \multicolumn{1}{c}{(internal)} & 1 \\
Kernel computation & \multicolumn{1}{c}{(internal)} & 2 \\
Whitening filter & 1 (WOLA) & 1 (AFIR) \\
Drift correction & 0 & 1 (AFIR) \\
\hline
Total whitening elements & 1 & 5 \\
\end{tabular}
\end{ruledtabular}
\end{table}

For multi-detector operation (H1 + L1 + V1), each detector receives its own independent whitening chain (Spectrum $\to$ WhiteningKernel $\to$ AFIR\textsubscript{whiten} $\to$ DriftKernel $\to$ AFIR\textsubscript{drift}).
The LLOID SVD template bank and coincidence logic downstream are shared across detectors~\cite{cannon_toward_2012, huang_sgnl_2025}.
The per-detector whitening chains are independent and can operate at different PSD update rates if the detectors exhibit different drift characteristics.

\section{Detailed Latency Accounting}
\label{app:latency-accounting}

This appendix provides the full itemized latency budget for both the LP baseline and the GTSP causal architecture, explains the stride absorption mechanism for the adjoint correction kernel, and discusses the computational cost tradeoffs.
All latency measurements are taken from the production \textsc{sgnl} pipeline running in wall-clock-throttled real-time mode, with latency defined as the data age at each processing stage: $\tau = t_{\mathrm{wall}} - t_{\mathrm{GPS}}$.

\subsection{LP latency budget (WOLA whitening)}

The LP pipeline's latency comprises four components, all scaling with the PSD FFT length $T$:

\begin{table}[h]
\caption{Itemized latency budget for the LP (WOLA) pipeline as a function of PSD estimation window length $T$.}
\label{tab:lp-latency-detail}
\begin{ruledtabular}
\begin{tabular}{lcccc}
Component & Formula & $T{=}4$\,s & $T{=}8$\,s & $T{=}16$\,s \\
\hline
Source stride & 1.0\,s & 1.0 & 1.0 & 1.0 \\
Look-ahead & $T/4$ & 1.0 & 2.0 & 4.0 \\
Zero-padding & $T/4$ & 1.0 & 2.0 & 4.0 \\
Stride mismatch & $\max(0, T/4{-}1)$ & 0.0 & 1.0 & 3.0 \\
\hline
\textbf{Total} & Eq.~\eqref{eq:latency-lp} & \textbf{3.0} & \textbf{6.0} & \textbf{12.0} \\
\end{tabular}
\end{ruledtabular}
\end{table}

The \textit{source stride} (1.0\,s) is the data delivery cadence, i.e., the irreducible delay between data acquisition and the first opportunity for the pipeline to process it.
The \textit{look-ahead} ($T/4$) arises because the symmetric WOLA spectral whitening has an effective group delay of $N/2$, requiring $N/2$ samples of future data at every stride boundary.
The \textit{zero-padding} ($T/4$) is a GPS timestamp shift introduced by the WOLA's zero-pad buffer, which causes the output GPS label to reference data that is $T/4$ seconds older than the stride boundary.
The \textit{stride mismatch} penalty appears when $T > 4$\,s: the WOLA stride ($T/4$) exceeds the source stride (1.0\,s), forcing the pipeline to buffer additional source deliveries before it can fire.

To illustrate, consider the $T = 8$\,s case.
The WOLA requires $8192$ samples per stride with $4096$ samples of overlap from the previous stride.
At wall-clock time $t = 1$\,s, the source delivers GPS~0 (2048 samples); the WOLA adapter has 2048 of the needed 8192.
At $t = 4$\,s (GPS~3), the adapter reaches 8192 samples and fires, producing output labeled GPS~$-2$ (shifted backward by $z_{\mathrm{whiten}} = T/4 = 2.0$\,s).
This output is discarded as a startup transient.
The first valid output (GPS~0) does not emerge until $t = 6$\,s, giving a steady-state latency of 6.0\,s.

\subsection{GTSP latency budget (causal AFIR + drift correction)}

The GTSP pipeline's latency has only two components and is independent of $T$:

\begin{table}[h]
\caption{Itemized latency budget for the GTSP (causal AFIR) pipeline.
The total is constant at 2.0\,s for all PSD lengths, provided $L \leq 2048$ taps.}
\label{tab:mp-latency-detail}
\begin{ruledtabular}
\begin{tabular}{lcc}
Component & Formula & Value \\
\hline
Source stride & 1.0\,s & 1.0 \\
Drift AFIR stride & 1.0\,s & 1.0 \\
Anti-causal delay & absorbed & 0.0 \\
\hline
\textbf{Total} & Eq.~\eqref{eq:latency-mp} & \textbf{2.0} \\
\end{tabular}
\end{ruledtabular}
\end{table}

The whitening AFIR contributes only $\tau_{\mathrm{s}}$ because the minimum-phase filter is strictly causal: its impulse response vanishes for $t < 0$, so the AFIR requires only past data from its overlap buffer to produce each output sample.
The overlap buffer ($N_w - 1$ samples, where $N_w = T_{\mathrm{PSD}} \times f_s$) is filled once at startup and then slides forward by one stride per firing.
After startup, the AFIR needs only one stride of new data to fire, regardless of the kernel length.

The drift AFIR contributes a second $\tau_{\mathrm{s}}$ as the architectural cost of the adjoint correction stage.
It applies $\mathcal{K}^\dagger$ using $L - 1$ samples of look-ahead from its overlap buffer, which consists entirely of already-received data retained from previous strides, so no additional wall-clock waiting is required.
The anti-causal delay $(L{-}1)/f_s$ is encoded in the output GPS label but does not contribute to wall-clock latency provided $L \leq f_s \cdot \tau_{\mathrm{s}}$.

\subsection{Stride absorption: mechanism and limits}

The anti-causal delay $(L{-}1)/f_s$ from the adjoint correction kernel appears in the drift AFIR's measured data age but is absorbed by the downstream LLOID element's stride alignment for $L \leq 2048$ taps:

\begin{table}[h]
\caption{Measured latency at the drift AFIR output and the LLOID output as a function of kernel truncation length $L$.
For $L \leq 2048$, the anti-causal delay is fully absorbed.}
\label{tab:stride-absorption-detail}
\begin{ruledtabular}
\begin{tabular}{lcccc}
$L$ (taps) & $(L{-}1)/f_s$ (s) & Drift AFIR (s) & LLOID (s) & Absorbed? \\
\hline
128 & 0.062 & 1.08 & 2.03 & Yes \\
512 & 0.250 & 1.27 & 2.03 & Yes \\
2048 & 1.000 & 2.02 & 2.04 & Yes \\
4096 & 2.000 & 3.02 & 3.03 & No \\
8192 & 4.000 & 5.02 & 5.03 & No \\
\end{tabular}
\end{ruledtabular}
\end{table}

The breakpoint is $L = f_s \cdot \Delta t_{\mathrm{stride}} + 1 = 2049$ taps.
For $L \leq 2048$ (1.0\,s at 2048\,Hz), the anti-causal delay fits within one stride of buffered data and adds zero wall-clock latency to the LLOID output.
Above this threshold, the AFIR must wait for additional data, and the constant-latency property of Eq.~\eqref{eq:latency-mp} is lost.

\subsection{Computational cost}

The GTSP pipeline's main computational costs are the cepstral projection $\mathcal{P}_+$ used in both kernel computations and the per-stride AFIR correlations.
Let $N = T_{\mathrm{PSD}} \times f_s$ denote the PSD FFT length in samples.

The cepstral projection involves five operations: elementwise log-spectrum ($\mathcal{O}(N)$), inverse FFT to the cepstral domain ($\mathcal{O}(N \log N)$), cepstral liftering ($\mathcal{O}(N)$), forward FFT ($\mathcal{O}(N \log N)$), and elementwise exponentiation ($\mathcal{O}(N)$), for a total of $\mathcal{O}(N \log N)$ per kernel computation.
The LP WOLA's per-stride cost (FFT of the data chunk, pointwise spectral division, inverse FFT, overlap-add) is also $\mathcal{O}(N \log N)$.

The GTSP incurs an additional $\mathcal{O}(N \log N)$ for the drift correction kernel, but this computation fires only when the PSD changes by more than the similarity threshold (Appendix~\ref{app:pipeline-topology}), not at every stride.
The per-stride cost of the whitening AFIR is $\mathcal{O}(N \times \tau_{\mathrm{s}} f_s)$ in the time domain; the drift AFIR is significantly cheaper at $\mathcal{O}(L \times \tau_{\mathrm{s}} f_s)$ since $L \ll N$ (e.g., $L = 128$ vs.\ $N = 16384$).
The overall per-stride cost is comparable to the LP WOLA.

\section{Supplementary Validation of the Correction Kernel}
\label{app:unit-proof}

The statistical pipeline results of Section~\ref{sec:pipeline-skymap} establish that the drift correction preserves timing and phase accuracy across 15,347 BBH injections on real O3 data.
Here we complement those results with a controlled, single-injection experiment that isolates the effect of the correction kernel $\mathcal{K}^\dagger$ on the matched-filter peak shape under known PSD perturbations.

We inject a single BNS signal into Gaussian noise and apply the adjoint correction via direct time-domain cross-correlation (\texttt{scipy.signal.correlate}), bypassing all pipeline infrastructure.
Three perturbation levels are tested: 15\%, 30\%, and 50\% RMS deviation between $S_{\mathrm{ref}}$ and $S_{\mathrm{live}}$.
For each level, we compare the matched-filter output of three configurations: (i)~the \textit{optimal} filter, constructed from the true live PSD; (ii)~the \textit{drift-corrected} filter, obtained by applying $\mathcal{K}^\dagger$ to the reference-whitened template; and (iii)~the \textit{uncorrected} filter, which uses the reference whitening without correction.

\begin{figure*}[t]
    \centering
    \includegraphics[width=\textwidth]{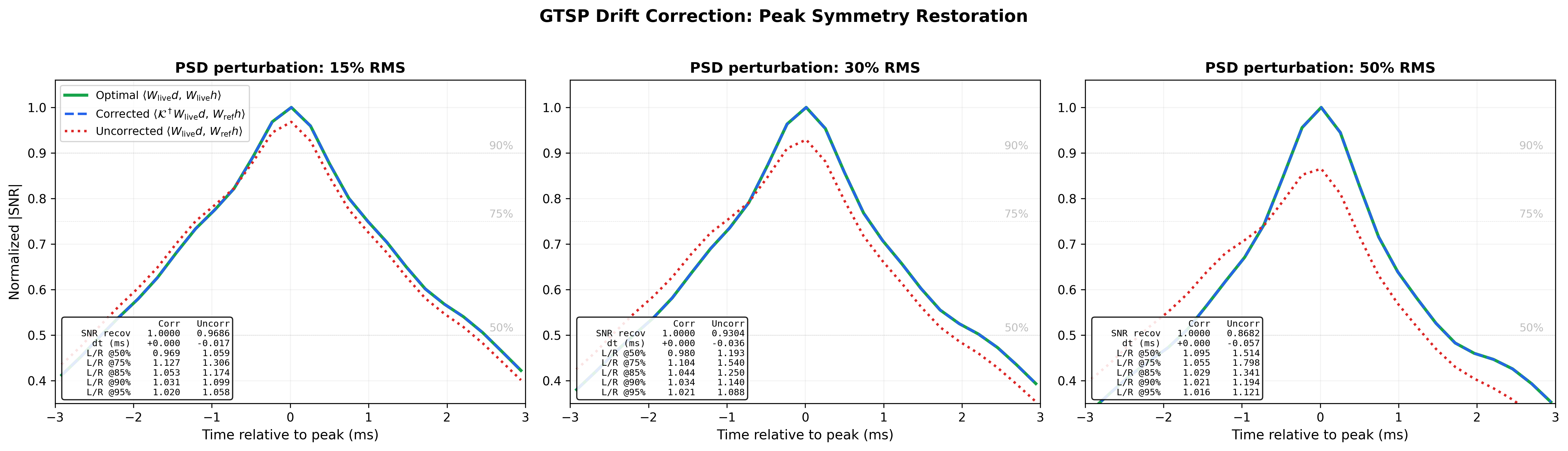}
    \caption{Unit-level demonstration of the holonomic correction kernel at three PSD perturbation levels (15\%, 30\%, 50\% RMS).
    The drift-corrected matched-filter peak (blue dashed) recovers the optimal SNR (green solid) to six significant figures, while the uncorrected peak (red dotted) loses up to 13\% and develops visible asymmetry.}
    \label{fig:unit-proof}
\end{figure*}

The results are summarized in Table~\ref{tab:unit-proof} and illustrated in Fig.~\ref{fig:unit-proof}.

\begin{table}[h]
\caption{Unit-level SNR recovery and timing accuracy under controlled PSD perturbation.
The corrected configuration recovers the optimal SNR to six significant figures with zero timing offset at all perturbation levels.}
\label{tab:unit-proof}
\begin{ruledtabular}
\begin{tabular}{lccc}
Perturbation & \multicolumn{2}{c}{SNR recovery} & Timing offset \\
 & Corrected & Uncorrected & Corrected (ms) \\
\hline
15\% RMS & 0.999997 & 0.969 & 0.000 \\
30\% RMS & 0.999994 & 0.930 & 0.000 \\
50\% RMS & 0.999988 & 0.868 & 0.000 \\
\end{tabular}
\end{ruledtabular}
\end{table}

The corrected matched-filter peak is indistinguishable from the optimal to six significant figures at all perturbation levels, with exactly zero timing offset.
The uncorrected peak loses up to 13\% of the optimal SNR at 50\% perturbation and develops visible asymmetry: the left-right width ratio at 50\% of peak height degrades from 1.00 (optimal) to 0.86 (uncorrected), reflecting the dispersive phase distortion predicted by the geometric framework~\cite{KenningtonBlack2026_GTSP1}.
The corrected peak restores perfect symmetry at all perturbation levels.

This controlled experiment confirms that the adjoint identity $\langle d, \mathcal{K}\psi_{\mathrm{ref}}\rangle = \langle \mathcal{K}^\dagger d, \psi_{\mathrm{ref}}\rangle$ is numerically exact, and that the correction kernel eliminates both the SNR loss and the peak asymmetry introduced by PSD mismatch.
The six-significant-figure agreement at 50\% perturbation, well beyond the drift amplitudes encountered in the O3 segment used for the pipeline validation, provides confidence that the framework will perform reliably under the more demanding non-stationarity of future observing runs.

\subsection{Residual distributions from the O3 injection campaign}

\begin{figure*}[t]
    \centering
    \includegraphics[width=\textwidth]{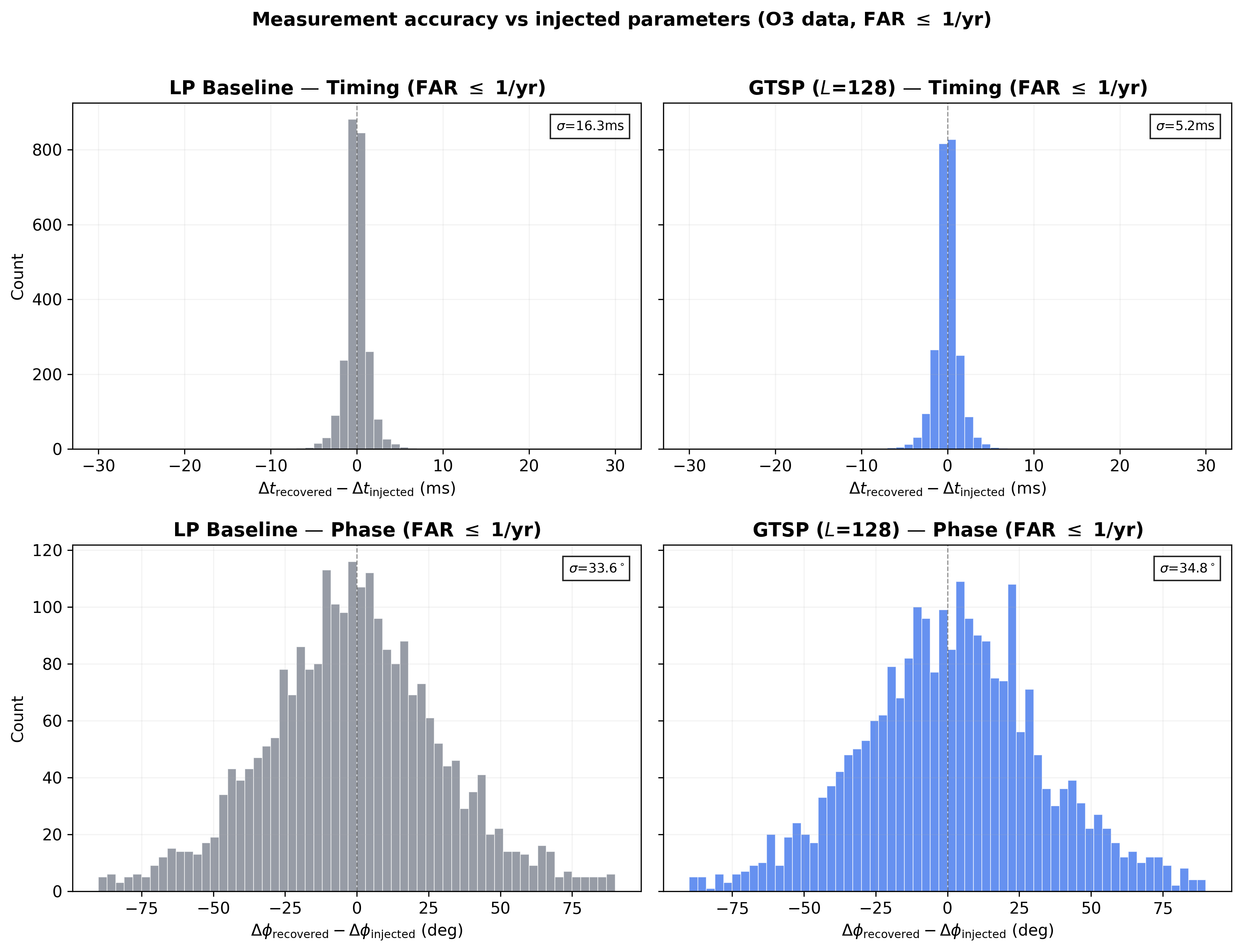}
    \caption{Residual distributions of inter-detector timing (top) and phase (bottom) relative to injected truth values on O3 data, for LP (left) and drift-corrected GTSP $L{=}128$ (right) at FAR~$\leq 1\,\mathrm{yr}^{-1}$.}
    \label{fig:residual-distributions}
\end{figure*}

Figure~\ref{fig:residual-distributions} shows the residual distributions of inter-detector timing and phase relative to injected truth from the O3 pipeline campaign described in Section~\ref{sec:pipeline-skymap}.
The GTSP architecture achieves notably better timing precision relative to injected truth, consistent with the causal filter's preservation of the signal's intrinsic group delay.
Phase precision is comparable between the two pipelines, with similar tail structure and no systematic bias.

\bibliography{references}

\end{document}